\documentclass[journal,draftcls,onecolumn,12pt,twoside]{IEEEtranTCOM}
\usepackage{graphicx}
\usepackage{amsmath, amsfonts, amssymb, amsthm}
\usepackage{enumerate}
\usepackage{multirow}
\usepackage{color}
\theoremstyle{definition} \newtheorem{theorem}{Theorem}
\theoremstyle{definition} \newtheorem{corollary}[theorem]{Corollary}
\theoremstyle{definition} \newtheorem{proposition}[theorem]{Proposition}
\theoremstyle{definition} 
\theoremstyle{definition} \newtheorem{lemma}[theorem]{Lemma}
\theoremstyle{definition} \newtheorem{algorithm}{Algorithm}

\theoremstyle{definition} 
\theoremstyle{definition} \newtheorem*{remark}{Remark}
\theoremstyle{definition} 
{\end{list}}

\begin{document}
%

\sloppy

\title{On Vector Linear Solvability of Multicast Networks}
\author{\IEEEauthorblockN{Qifu~Tyler~Sun,~Xiaolong Yang,~Keping~Long,~Xunrui~Yin,~and~Zongpeng~Li}
\thanks{The preliminary results of this paper were partially presented at 2015 IEEE International Conference on Communications and at 2015 IEEE International Symposium on Information Theory.}
}
\maketitle

\begin{abstract}
Vector linear network coding (LNC) is a generalization of the conventional scalar LNC, such that the data unit transmitted on every edge is an $L$-dimensional vector of data symbols over a base field GF($q$). Vector LNC enriches the choices of coding operations at intermediate nodes, and there is a popular conjecture on the benefit of vector LNC over scalar LNC in terms of alphabet size of data units: there exist (single-source) multicast networks that are vector linearly solvable of dimension $L$ over GF($q$) but not scalar linearly solvable over any field of size $q' \leq q^L$. This paper introduces a systematic way to construct such multicast networks, and subsequently establish explicit instances to affirm the positive answer of this conjecture for \emph{infinitely many} alphabet sizes $p^L$ with respect to an \emph{arbitrary} prime $p$. On the other hand, this paper also presents explicit instances with the special property that they do not have a vector linear solution of dimension $L$ over GF(2) but have scalar linear solutions over GF($q'$) for some $q' < 2^L$, where $q'$ can be odd or even. This discovery also unveils that over a given base field, a multicast network that has a vector linear solution of dimension $L$ does not necessarily have a vector linear solution of dimension $L' > L$.

\end{abstract}
\begin{keywords}
Vector network coding, scalar network coding, multicast networks, alphabet size, direct sum.
\end{keywords}

\section{Introduction}
In the conventional theory of linear network coding (LNC) \cite{LiYeungCai03}\cite{KoetterMedard03}, the data unit transmitted along every edge of unit capacity in a network consists of a single data symbol belonging to a base field GF($q$). Every outgoing edge of a node $v$ transmits a data symbol that is a GF($q$)-linear combination of the incoming data symbols to $v$. Such a coding mechanism is referred to as \emph{scalar LNC}.

\begin{figure}[!htbp]
\centering
\scalebox{1}
{\includegraphics{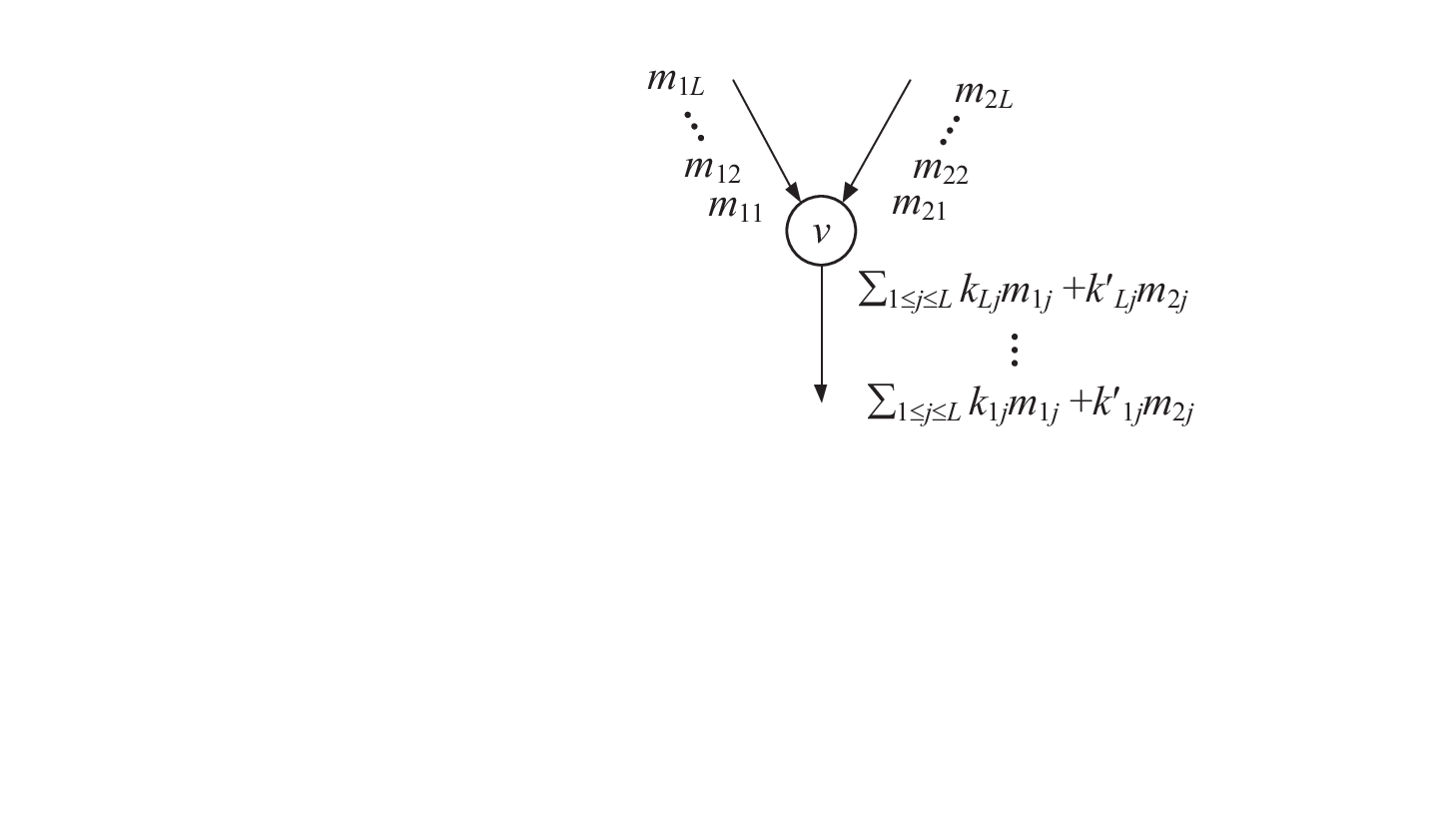}}
\caption{Consider an intermediate node $v$ with $1$ outgoing edge and $2$ incoming edges in a network. In vector LNC, $v$ receives two vectors $(m_{11}, \cdots, m_{1L})$ and $(m_{21}, \cdots, m_{2L})$ of $L$ data symbols belonging to a base field GF($q$). The data unit transmitted on the outgoing edge of $v$ is also a vector of $L$ data symbols over GF($q$), in which each data symbol is a GF($q$)-linear combination of all $2L$ incoming data symbols to $v$.}
\label{Fig:Coding_operation_vectorLNC}
\end{figure}

A generalization of scalar LNC is \emph{vector LNC} \cite{Medard03} or \emph{block LNC} \cite{Jaggi04_On_LNC}, which models the data unit transmitted along every edge of unit capacity as an $L$-dimensional vector of data symbols over a base field GF($q$), and concomitantly defines the coding operations performed at every intermediate node as GF($q$)-linear combinations of all data symbols in incoming data unit vectors, as illustrated in Fig.~\ref{Fig:Coding_operation_vectorLNC}.

The introduction of the concept of vector LNC stems from its potential to enrich the choices of coding operations at intermediate nodes in a network. The potential of vector LNC has been considered from several different aspects (See for example \cite{Medard03}-\cite{JiajiaGuo15}). In particular, the work in \cite{Medard03} demonstrated a classic multi-source multicast network which has a simple vector linear solution of dimension $2$ over GF(2) but does not have a scalar linear solution over any base field. It was also noted in \cite{Medard03} that the network constructed in \cite{Lehman03} which is not scalar linearly solvable over any field has a vector linear solution. There is another exemplifying network proposed in \cite{Riis} which does not have a scalar linear solution over any field but has a vector linear solution of dimension $3$ over GF(2). These exemplifying networks manifest the superiority of vector LNC over scalar LNC in terms of enabling a linear solution.

Up to now, most studies on vector LNC have been in the context of general (non-multicast) networks. Specific to a (single-source) multicast network, though it is well known that there is a scalar linear solution over a field with size no smaller than the number of receivers \cite{Jaggi05}, there are still benefits to consider vector LNC, as summarized in \cite{Fragouli11}. In particular, the alphabet size of data units is a key factor that affects the linear solvability of a multicast network. Under the same alphabet size $q^L$, in which case the transmission delay of a data unit along an edge is same, vector LNC of dimension $L$ over GF($q$) provides much more choices for coding operations than scalar LNC over GF($q^L$), and every scalar linear code over GF($q^L$) can be transformed into a vector linear code of dimension $L$ over GF($q$), so that the scalar linear code is a solution if and only if its corresponding vector linear code is a solution too. Thus, a network has a scalar linear solution over GF($q^L$) \emph{only if} it has a vector linear solution of dimension $L$ over GF($q$). It would be natural to conceive the following benefit of vecor LNC, as conjectured in \cite{Fragouli11}:
\begin{itemize}
\item There exists a multicast network that is vector linearly solvable of dimension $L$ over GF($q$), but not scalar linearly solvable over GF($q'$) for any $q' \leq q^L$.
\end{itemize}
If proven true, this conjecture will imply practical benefit of vector LNC in terms of reducing the alphabet size to yield a solution on a multicast network, which is a fundamental research topic in the network coding literature. However, even though the work in \cite{Fragouli11} indicated the possible correctness of this conjecture from the perspective of multivariate determinant polynomials of transfer matrices, it failed to provide explicit multicast networks to verify its correctness.

In the first part of the paper, we propose a systematic way to construct a multicast network vector linearly solvable over GF($q$) at dimension $L$ but not scalar linearly solvable over GF($q'$) for any $q' \leq q^L$. Explicit multicast networks are also subsequently constructed based on this method so that the aforementioned conjecture is proven to be correct for \emph{infinitely many} alphabet sizes $p^L$ with respect to an \emph{arbitrary} prime $p$. Moreover, some of the illustrated multicast networks vector linearly solvable at dimension $L$ over GF($p$) do not have a scalar linear solution over GF($q'$), not only for all those $q' \leq p^L$, but also for some $q' > p^L$. This affirms that in some multicast networks, vector LNC can indeed be superior to scalar LNC, in a stronger sense than as conjectured, in terms of alphabet size of data units to yield a solution. %
The vector coding techniques we propose to beat scalar codes have several implications:
\begin{itemize}
\item Scalar linear solutions over respective alphabets $\mathrm{GF}(q^{L_1}), \cdots, \mathrm{GF}(q^{L_m})$ do not necessarily imply another scalar linear solution over $\mathrm{GF}(q^{L_1 + \cdots + L_m})$, but they guarantee a vector linear solution of dimension $L_1 + \cdots + L_m$ over GF($q$).
\item For scalar linear codes over respective alphabets $\mathrm{GF}(q^{L_1}), \cdots, \mathrm{GF}(q^{L_m})$, even in the case that none of them has a solution, it is still possible to combine their corresponding vector linear codes, by direct sum, to form a vector linear solution of dimension $L_1 + \cdots + L_m$ over GF($q$).
\end{itemize}

\vspace{3pt}

In the second part of the paper, we compare the alphabet size requirements for scalar and vector linear solvability of multicast networks from another direction. Specifically, now that the non-existence of a vector linear solution of dimension $L$ over GF($q$) implies the non-existence of a scalar linear solution over GF($q^L$), a natural question is whether it can further imply the non-existence of a scalar linear solution over every GF($q'$) with $q' \leq q^L$. At a first glance, one might be inclined to believe its correctness. However, as we shall demonstrate, the answer to this question is \emph{negative}. Another contribution of this paper is to show explicit multicast networks, for the first time in the literature, which do not have a vector linear solution of dimension $L$ over GF($2$) but have a scalar linear solution over GF($q'$) for some $q' < 2^L$, where $q'$ can be odd or a power of $2$. This discovery suggests that it is also possible for scalar LNC to outperform vector LNC (of dimension $L \geq 2$) in multicast networks, in terms of using a smaller alphabet to yield a solution. More importantly, it further discloses that
\begin{itemize}
\item over a given base field, a multicast network vector linearly solvable of dimension $L$ is not necessarily vector linearly solvable of dimensions $L'$ with $L' > L$.
\end{itemize}
This discovery is intriguing in the sense that it appears to contradict the folklore on multicast networks: the larger the alphabet block length, the more likely a linear solution exists.

Recently, a few multicast networks were discovered in \cite{Sun_TIT} with the intriguing property that they are scalar linearly solvable over a small field but not necessarily scalar linearly solvable over a larger field. They share a common topological structure, and can thus be subsumed in a particular class of multicast networks, whose scalar linear solvability is completely characterized in \cite{Sun_15}. One of the fundamental building blocks for the results obtained in this paper is the further analysis of the vector linear solvability of this special class of multicast networks, which was not dealt with in \cite{Sun_TIT} and \cite{Sun_15}.

The remainder of this paper is organized as follows. In Section II, we establish the mathematical notations to be adopted and review some useful fundamental results of vector and scalar LNC. In Section III, we present a general way to construct multicast networks vector linearly solvable over GF($q$) of dimension $L$ but not scalar linearly solvable over GF($q'$) for any $q' \leq q^L$, and present instances for an arbitrary prime $p$ and infinitely many alphabet sizes $p^L$. In Section IV, we verify that on multicast networks smaller alphabets can be better than larger ones for yielding a vector linear solution. Section V summarizes the paper.

\vspace{5pt}

\section{Preliminaries}
\subsection{Mathematical Model for Vector Linear Codes}
This work focuses on a single-source multicast network, which is modeled as a finite directed acyclic multigraph, with a unique source node $s$ and a set $T$ of receivers. For a node $v$ in the network, denote by $In(v)$ and $Out(v)$, respectively, the set of its incoming and outgoing edges. Every edge has unit capacity and every outgoing edge from the source $s$ transmits a data unit generated by $s$. Write $|Out(s)| = \omega$, which will be referred to as the source dimension of the network. Then there are totally $\omega$ source data units to be transmitted across the network. A topological order is assumed on the set of edges led by ones in $Out(s)$. For every receiver $t \in T$, based on the data units received from edges in $In(t)$, its goal is to recover the source data units generated from $s$. The maximum flow for every receiver $t \in |T|$, which is defined to be the maximum number of edge-disjoint paths leading from $s$ to $t$, is assumed to be $\omega$.

In the conventional scalar LNC, the data unit transmitted along every edge $e$ merely consists of a single data symbol belonging to a symbol alphabet which is mathematically modeled as a finite field GF($q$). A \emph{scalar linear code} is an assignment of a \emph{local encoding kernel} $k_{d,e} \in \mathrm{GF}(q)$ to every pair $(d, e)$ of edges such that $k_{d,e} = 0$ when $(d, e)$ is not an adjacent pair of edges. Every scalar linear code uniquely determines a global encoding kernel $\mathbf{f}_e$, which is an $\omega$-dimensional column vector over GF($q$), for each edge $e$ in the network. On a multicast network, a scalar linear code is called a \emph{scalar linear solution} if for every receiver $t \in T$, the juxtaposition $[\mathbf{f}_e]_{e\in In(t)}$ of the global encoding kernels for edges incoming to $t$ has full rank $\omega$.

As a generalization of scalar LNC, vector LNC models the data unit transmitted along every edge $e$ to be an $L$-dimensional \emph{row vector} $\mathbf{m}_e$ of data symbols over a base field GF($q$). Thus, the mathematical model for the data unit alphabet in vector LNC is a vector space GF($q$)$^L$ rather than a finite field. Under the new mathematical structure of data units, the model of scalar LNC can be naturally extended to vector LNC as follows.

A \emph{vector linear code of dimension $L$ over} GF($q$), or \emph{a vector linear code over} GF($q$)$^L$ for short, is an assignment of a \emph{local encoding kernel}  $\mathbf{K}_{d,e}$, which is an $L\times L$ matrix over $\mathrm{GF}(q)$, to every pair $(d, e)$ of edges such that $\mathbf{K}_{d,e}$ is the zero matrix $\mathbf{0}$ when $(d, e)$ is not an adjacent pair. Then, for every edge $e$ outgoing from a non-source node $v$, the data unit vector of data symbols transmitted on $e$ is
\[
\mathbf{m}_e = \sum_{d\in In(v)} \mathbf{m}_d\mathbf{K}_{d,e}
\]
Furthermore, every vector linear code uniquely determines a global encoding kernel $\mathbf{F}_{e}$, which is an $\omega L\times L$ matrix over GF($q$), for every edge $e$ such that
\begin{itemize}
\vspace{3pt}
\item The columnwise juxtaposition $[\mathbf{F}_e]_{e\in Out(s)}$ of $\mathbf{F}_e$ for $e\in Out(s)$ forms an $\omega L\times\omega L$ identity matrix $\mathbf{I}$;
\vspace{3pt}
\item For every outgoing edge $e$ from a non-source node $v$, $\mathbf{F}_e = \sum_{d\in In(v)} \mathbf{F}_d\mathbf{K}_{d,e}$.
\end{itemize}
\vspace{3pt}
Correspondingly, the data unit vector transmitted along every edge $e$ can also be represented as
\[
\mathbf{m}_e = [\mathbf{m_d}]_{d\in Out(s)} \mathbf{F}_e.
\]
A vector linear code over GF($q$)$^L$ is called a \emph{vector linear solution} if for every receiver $t \in T$, the juxtaposition $[\mathbf{F}_e]_{e\in In(t)}$ of the global encoding kernels for edges incoming to $t$ has full rank $\omega L$. Correspondingly, there is an $L|In(t)|\times L|Out(s)|$ decoding matrix $\mathbf{D}_t$ over GF($q$) for every receiver $t$ such that
the source data units can be recovered at $t$ via
\begin{align*}
[\mathbf{m}_e]_{e\in In(t)}\mathbf{D}_t &= \left([\mathbf{m_d}]_{d\in Out(s)} [\mathbf{F}_e]_{e\in In(t)} \right) \mathbf{D}_t \\
&=[\mathbf{m_d}]_{d\in Out(s)} \left([\mathbf{F}_e]_{e\in In(t)} \mathbf{D}_t\right) \\
&= [\mathbf{m_d}]_{d\in Out(s)} \mathbf{I} = [\mathbf{m_d}]_{d\in Out(s)}.
\end{align*}

A scalar linear code can be regarded as a vector linear code from two different facets. On one hand, it is straightforward to see that every scalar linear code over GF($q^L$) is naturally a vector linear code of dimension 1 over GF($q^L$). On the other hand, let $\Phi$ be a mapping from GF($q^L$) into the ring of $L \times L$ matrices over GF($q$) via
\begin{equation}
\label{eqn:Matrix_Representation_Homomorphism}
\begin{matrix}
\Phi(0) = \mathbf{0}, \\
\Phi(\gamma^k) = \mathbf{C}_p^k,~0 \leq k \leq q^L - 2,
\end{matrix}
\end{equation}
where $\mathbf{C}_p$ is the $L\times L$ companion matrix of a primitive polynomial $p(x)$ of degree $L$ over GF($q$), and $\gamma$ is a fixed root of $p(x)$, that is, a primitive element of GF($q^L$). As a consequence of the Cayley-Hamilton theorem, this mapping is a homomorphism, and $\{\mathbf{0}, \mathbf{C}^0, \cdots, \mathbf{C}^{q^L-2}\}$ forms a matrix representation of the finite field GF($q^L$) (See, for example, \cite{Wardlaw}). Then, every scalar linear code over GF($q^L$) with local encoding kernels $(k_{d,e})$ corresponds to a vector linear code over GF($q$)$^L$ with local encoding kernels prescribed as
\begin{equation}
\mathbf{K}_{d,e} = \Phi(k_{d,e}),
\end{equation}
and moreover, based on the homomorphic property of $\Phi$, we can derive the following result.

\begin{proposition}
\label{prop:matrix_representation_scalar_solution}
Given a (not necessarily multicast) network, a scalar linear code over GF($q^L$) with local encoding kernels $(k_{d,e})$ is a solution if and only if the corresponding vector linear code over $\mathrm{GF}(q)^L$ with local encoding kernels $\mathbf{K}_{d,e} = \Phi(k_{d,e})$ qualifies as a solution too. Moreover, if $\mathbf{D}_t$ is a decoding matrix of the scalar linear solution for receiver $t$, then $\Phi(\mathbf{D}_t)$ is also a decoding matrix of the corresponding vector linear solution for $t$, where $\Phi$ is applied componentwise for the entries in $\mathbf{D}_t$.
\end{proposition}

In summary, Table \ref{table:scalar_vs_vector_LNC} compares the mathematical structures of scalar and vector LNC.

\begin{table}[!htbp]
\caption{Comparison of scalar and vector LNC when the alphabet size of data units is fixed to $q^L$ with $q$ equal to a prime power}
\label{table:scalar_vs_vector_LNC}
\begin{center}
\begin{tabular}{c|c|c}
  \hline
     & \multirow{2}{*}{Scalar LNC} & \multirow{2}{*}{Vector LNC} \\
     & & \\ \hline
  \multirow{2}{*}{Data unit alphabet} & \multirow{2}{*}{Base field GF($q^L$)} & \multirow{2}{*}{Vector space GF($q$)$^L$}\\
   & &  \\ \hline
  \multirow{2}{*}{Local encoding kernel} & \multirow{2}{*}{Element in GF($q^L$)} &  \multirow{2}{*}{$L\times L$ matrix over GF($q$)} \\
  & & \\ \hline
  \# of candidates for & \multirow{3}{*}{$q^L$} & \multirow{3}{*}{$q^{L^2}$} \\
  local encoding kernels & & \\
  (for adjacent pairs of edges) & & \\ \hline
\end{tabular}
\end{center}
\end{table}

\vspace{5 pt}

\subsection{A Special Class of Multicast Networks}
\begin{figure}[t]
\centering
\scalebox{0.58}
{\includegraphics{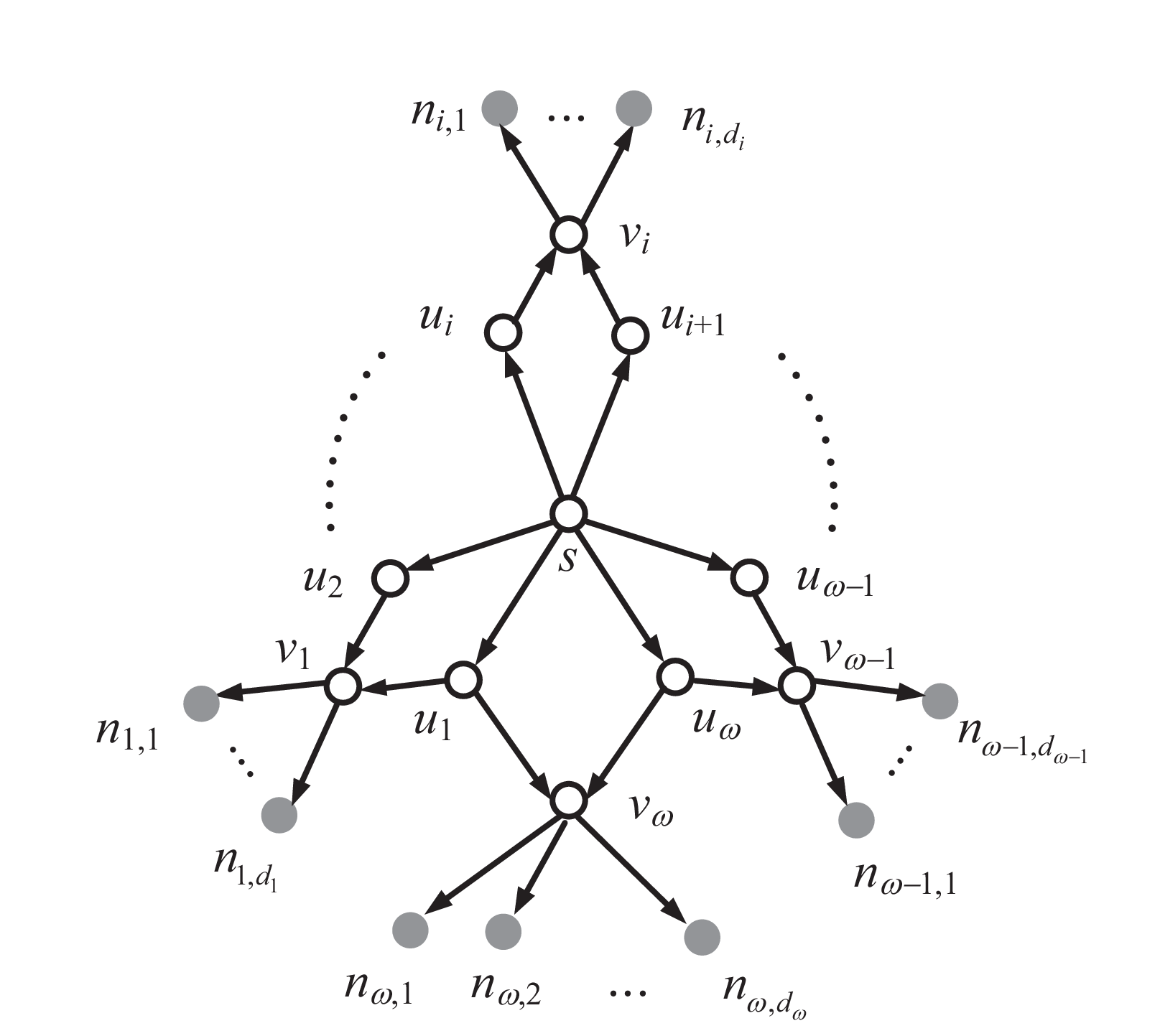}}
\caption{The network $\mathcal{N}_{\omega, \mathbf{d}}$ consists of nodes on 5 layers. Layer-1 consists of the source node $s$ only, and layer-4 nodes are depicted in grey. There is a non-depicted bottom-layer node connected from every set $N$ of $\omega$ layer-4 nodes with $maxflow(N) = \omega$. All bottom-layer nodes are receivers.}
\label{Fig:General_network_for_scalar}
\end{figure}

Recently, the first few known multicast networks that are scalar linearly solvable over GF($q$) but not necessarily over a larger GF($q'$) were discovered in \cite{Sun_TIT}. They share a similar topology and can be generalized into a class of multicast networks $\mathcal{N}_{\omega,\mathbf{d}}$, as replotted in Fig. \ref{Fig:General_network_for_scalar}, with topological parameters $\omega$ and $\mathbf{d} = (d_1, d_2, \cdots, d_\omega)$. The network $\mathcal{N}_{\omega,\mathbf{d}}$ has source dimension $\omega$, and consists of nodes on five layers. The source $s$ is the unique node in the first layer. There are $\omega$ layer-2 nodes $u_j$, $1 \leq j \leq \omega$, each of which is connected from $s$ by an edge. There are $\omega$ layer-3 nodes $v_j$, $1 \leq j \leq \omega$, each of which is connected from two upstream layer-2 nodes $u_j$ and $u_{j+1}$ ($u_{\omega+1}$ represents $u_1$) by a respective edge. For each layer-3 node $v_j$, there are $d_j > 1$ outgoing edges, each of which leads to a different downstream layer-4 (grey) node. Thus, the $\omega$-tuple $\mathbf{d} = (d_1, \cdots, d_\omega)$ controls the number of layer-4 nodes. There is a non-depicted bottom-layer node connected from every set $N$ of $\omega$ layer-4 nodes with $maxflow(N) = \omega$, that is, with $\omega$ edge-disjoint paths starting from $s$ and ending at nodes in $N$. All bottom-layer nodes are receivers.

The following is a concise formula for the scalar linear solvability of $\mathcal{N}_{\omega,\mathbf{d}}$ derived in \cite{Sun_15}.

\begin{theorem}
\label{thm:Concise_Scalar_Linear_Solvability_Characterization}
Consider a network $\mathcal{N}_{\omega,\mathbf{d}}$ with parameters $\omega$ and $\mathbf{d} = (d_1, d_2, \cdots, d_\omega)$. It is linearly solvable over GF($q$) if and only if there is positive divisor $d$ of $q - 1$ subject to
\begin{equation}
\label{eqn:equivalence_subgroup_order_perspective}
q \geq d\left(\left\lceil\frac{d_1}{d}\right\rceil + \cdots + \left\lceil\frac{d_\omega}{d}\right\rceil - \omega + 1 \right) + 2
\end{equation}
\end{theorem}

\begin{corollary}
\label{cor:Swirl_Network_Scalar_Solv}
The network $\mathcal{N}_{\omega,\mathbf{d}}$ with parameters $\omega$ and $\mathbf{d} = (2, 2, \cdots, 2)$ is called the \emph{Swirl Network} \cite{Sun_TIT}. As a consequence of Theorem \ref{thm:Concise_Scalar_Linear_Solvability_Characterization}, it is scalar linearly solvable over GF($q$) if and only if $q > \omega + 2$ or $q - 1$ is not a prime.
\end{corollary}

The analysis of the vector linear solvability of $\mathcal{N}_{\omega, \mathbf{d}}$ will be one of the building blocks for the main discoveries of this paper.

\section{Multicast Networks with Vector LNC Superior to Scalar LNC}
\begin{figure}[h]
\centering
\scalebox{.78}
{\includegraphics{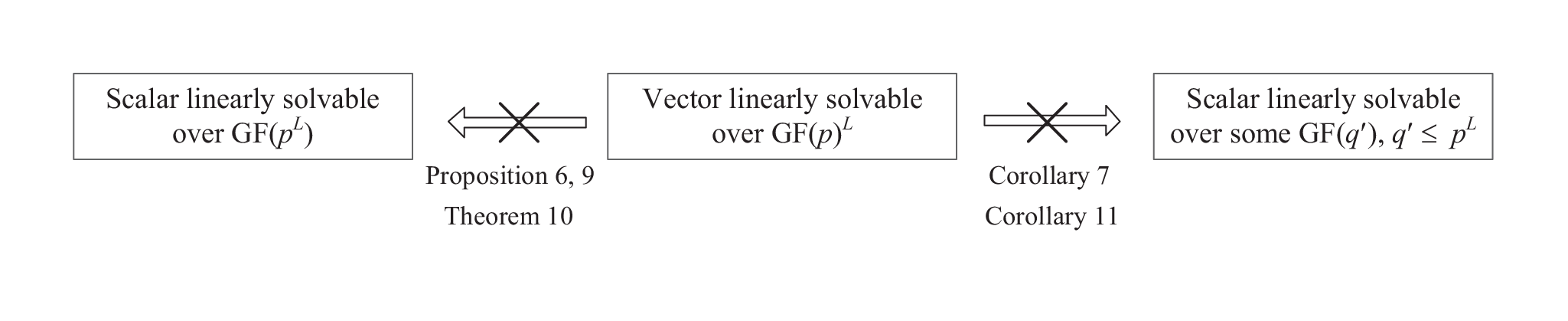}}
\caption{The main results to be established in Section III. Herein $p$ is an arbitrary prime.}
\label{Fig:Relation_Diagram_Part1}
\end{figure}

In this section, we shall first introduce a general method to construct multicast networks vector linearly solvable over GF($q$)$^L$ but not scalar linearly solvable over any GF($q'$) with $q' \leq q^L$. Then, we make use of this method to design infinitely many instances to verify that vector LNC can indeed outperform scalar LNC for multicast networks in terms of the required alphabet size to yield a solution. The main results to be established in this section are outlined in Fig. \ref{Fig:Relation_Diagram_Part1}.

\subsection{A General Construction Method}
Under the same alphabet size of data units, which is considered to be a prime power throughout the paper, the number of candidates to which the local encoding kernels can be assigned increases \emph{exponentially} from $q^L$ to $q^{L^2}$. Consequently, it is natural to conceive that vector LNC outperforms scalar LNC on a multicast network in the sense that the minimum alphabet size to yield a vector linear solution might be smaller than the minimum required in a scalar solution. However, to the best of our knowledge, no explicit demonstration of this advantage for vector LNC on multicast networks has ever been given, and this advantage was only partially confirmed in \cite{Fragouli11}. %
In the work of \cite{Fragouli11}, an algebraic framework is established to characterize the vector linear solvability of a multicast network, which can be regarded as a generalization of the classic algebraic framework in \cite{KoetterMedard03} that concentrates on scalar linear solvability. %
Specifically, the framework associates every receiver in a multicast network with a transfer matrix whose entries are multivariate polynomials. Correspondingly, it associates a multicast network with a multivariate polynomial obtained by the product of the determinants of all transfer matrices. It is then shown that a multicast network is vector linearly solvable over GF($q$)$^L$ if and only if there is an assignment of $L \times L$ matrices over GF($q$) to the variables in the associated polynomial under which the evaluation of this polynomial is an invertible matrix over GF($q$). %
Meanwhile, a multivariate polynomial was discovered in \cite{Fragouli11} which does not have such an assignment over GF($q$) for any $q \leq 2^{10}$, but has a feasible assignment over GF($2$)$^{10}$. However, that work did not show the existence of a multicast network that can be associated with this particular polynomial, and hence whether there exists a multicast network with the desired advantage of vector LNC remains elusive.

We next propose a general construction method, based on which the design of a multicast network vector linearly solvable over GF($q$)$^L$ but not scalar linearly solvable over GF($q'$) for every $q' \leq q^L$ reduces to the design of a multicast network vector linearly solvable over GF($q$)$^L$ but not scalar linearly solvable over GF($q^L$).

\begin{algorithm}
Let $\mathcal{N}_1$ be a multicast network with source dimension $\omega$ that is vector linearly solvable over $\mathrm{GF}(q)^L$ but not scalar linearly solvable over GF($q^L$). Set $n = q^L$. Construct a multicast network $\mathcal{N}$ of source dimension $\omega$ as follows:
\begin{itemize}
\item Create the unique source node $s'$ and another node $s$, as well as $\omega$ edges starting from $s'$ and ending at $s$.
\item Add $\mathcal{N}_1$ as a subnetwork of $\mathcal{N}$. Create $\omega$ edges from $s$ to the original source node $s_1$ of $\mathcal{N}_1$.
\item Add an $(n+1, 2)$-combination network $\mathcal{N}_2$ (See, e.g., \cite{Ngai_Yeung}\cite{Xiao07}), as depicted in Fig. \ref{Fig:Combination_Network}, to be another subnetwork of $\mathcal{N}$. Create 2 edges from $s$ to the original source node $s_2$ of $\mathcal{N}_2$.
\item For every original receiver $t$ of $\mathcal{N}_2$, create $\omega - 2$ edges from $s$ to $t$.
\end{itemize}
In this way, every node that is originally a receiver in $\mathcal{N}_1$ or $\mathcal{N}_2$ is also a receiver in $\mathcal{N}$.    \hfill $\blacksquare$
\end{algorithm}

\begin{theorem}
\label{thm:general_construction}
Let $\mathcal{N}_1$ be a multicast network that is vector linearly solvable over $\mathrm{GF}(q)^L$ but not scalar linearly solvable over GF($q^L$). The network $\mathcal{N}$ constructed by Algorithm 1 with $n = q^L$ has a vector linear solution over GF($q$)$^L$. However, it is not scalar linearly solvable over GF($q'$) for any $q' \leq q^L$, and not vector linearly solvable over GF($q'$)$^{L'}$ for any $q'^{L'} < q^L$.
\begin{proof}
The network $\mathcal{N}$ is vector linearly solvable over GF($q'$)$^{L'}$ or scalar linearly solvable over GF($q'$) if and only if so are the subnetworks $\mathcal{N}_1$ and $\mathcal{N}_2$.

It is well known that the $(n+1, 2)$-combination network is scalar linearly solvable over GF($q'$) if and only if $q' \geq n$. In a similar argument to characterize its scalar linear solvability, one can deduce that an $(n+1, 2)$-combination network is vector linear solvable over GF($q'$)$^{L'}$ if and only if there are $L' \times L'$ invertible matrices $\mathbf{A}_1, \cdots, \mathbf{A}_{n-1}$ over GF($q'$) such that
\begin{equation}
\label{eqn:combination_network_vector_LNC}
rank(\mathbf{A}_i -\mathbf{A}_j) = L',~~\forall 1 \leq i < j \leq n-1,
\end{equation}
Thus, $\{\mathbf{A}_1, \cdots, \mathbf{A}_{n-1}\}$, together with $\mathbf{0}$ form an $L'$-dimensional rank-metric code of distance $L'$ over GF($q'$). According to the Singleton bound for the rank-metric codes (See \cite{Gabidulin85} for example), there are at most $q'^{L'(L' - L' +1)} = q'^{L'}$ codewords for such a rank-metric code. Thus, if there are $L' \times L'$ invertible matrices $\mathbf{A}_1, \cdots, \mathbf{A}_{n-1}$ over GF($q'$) subject to (\ref{eqn:combination_network_vector_LNC}), then $n - 1 \leq q'^{L'} - 1$, \emph{i.e.} $q'^{L'} \geq n$. On the other hand, when $q'^{L'} \geq n$, a scalar linear solution can be constructed for the $(n+1,2)$-combination network over GF($q'^{L'}$), which in turn induces a vector linear solution over GF($q'$)$^{L'}$ according to Proposition \ref{prop:matrix_representation_scalar_solution}. We can now conclude that an $(n+1, 2)$-combination network is vector linearly solvable over GF($q'$)$^{L'}$ if and only if $q'^{L'} \geq n$.

In consequence, the subnetwork $\mathcal{N}_2$ of $\mathcal{N}$ has a vector linear solution over GF($q$)$^L$, but neither a scalar nor a vector linear solution when the alphabet size of data units is smaller than $q^L$. On the other hand, the subnetwork $\mathcal{N}_1$ of $\mathcal{N}$ is vector linearly solvable over GF($q$)$^L$ but not scalar linearly solvable over GF($q^L$). We can see when the alphabet size is no greater than $q^L$, the network $\mathcal{N}$ does not have any scalar linear solution, and has a vector linear solution only over GF($q$)$^L$.
\end{proof}
\end{theorem}

\begin{figure}[t]
\centering
\scalebox{0.66}
{\includegraphics{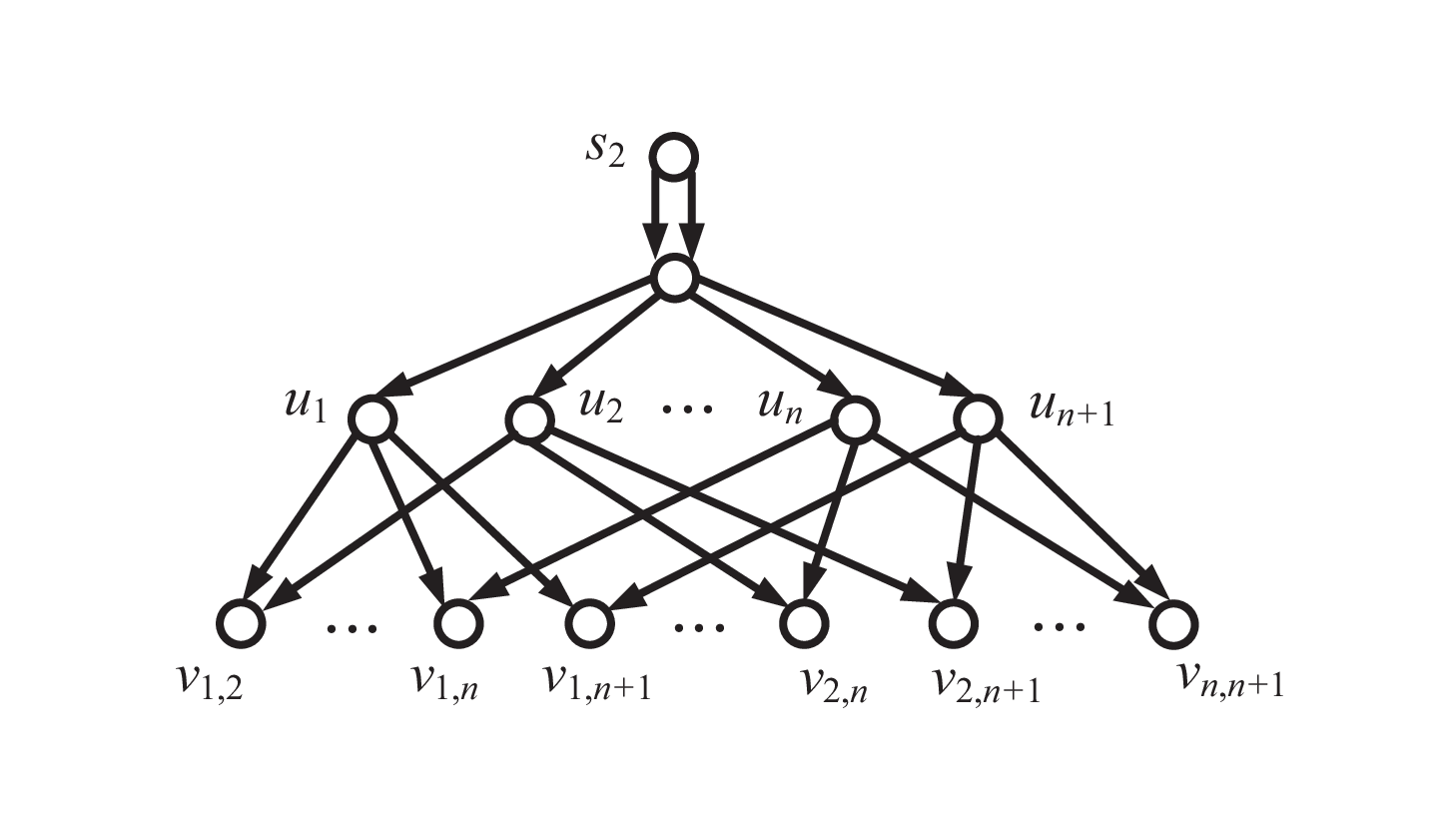}}
\caption{An $(n+1, 2)$-combination network $\mathcal{N}_2$. It is well known to be scalar linearly solvable over GF($q^L$) if and only if $q^L \geq n$. It can also be shown to be vector linearly solvable over GF($q$)$^L$ if and only if $q^L \geq n$.}
\label{Fig:Combination_Network}
\end{figure}

\vspace{5 pt}

\subsection{The First Explicit Network Construction}
In order to apply Algorithm 1 to construct a multicast network vector linearly solvable over $\mathrm{GF}(q)^L$ but not scalar linearly solvable over $\mathrm{GF}(q')$ for any $q' \leq q^L$, a key step that has not been explicated is to provide a multicast network that is vector linearly solvable over $\mathrm{GF}(q)^L$ but not scalar linearly solvable over $\mathrm{GF}(q^L)$. We next show that the Swirl Network with appropriate source dimension $\omega$ is actually the first known one with such a property.

Assume that the alphabet size of data units is $2^L$. When $2^L - 1$ is a prime, the Swirl Network with dimension $\omega \geq 2^L - 2$ does not have a scalar linear solution over GF($2^L$). Recall that a prime in the form of $2^L - 1$ is called a Mersenne prime. After examining the list of all known 48 Mersenne primes in the ascending order \cite{GIMPS}, we found that the $5^{th}$ one, $2^{13} - 1$, can be written as $2^4\cdot2^9 - 1$ but neither $2^4 - 1$ nor $2^9 - 1$ is a Mersenne prime. Thus, the Swirl Network turns out to be the first exemplifying multicast network scalar linearly solvable over both GF($q^{L_1}$) and GF($q^{L_2}$) but not over GF($q^{L_1+L_2}$). This has been noticed in \cite{Sun_TIT}.

Now consider a (possibly non-multicast) network and a scalar linear solution of it, with local encoding kernels denoted by $(k_{d,e,j})$, over GF($q^{L_j}$) for all $1 \leq j \leq m$. We can define a vector linear code of dimension $L := L_1 + L_2 + \cdots + L_m$ over GF($q$) with local encoding kernels prescribed by
\begin{equation}
\label{eqn:GEK_Scalar_to_Vector_Multi}
\mathbf{K}_{d,e} = \left[\begin{matrix}
\Phi(k_{d,e,1}) & \mathbf{0} & \cdots & \mathbf{0} \\
\mathbf{0} & \Phi(k_{d,e,2}) & \cdots & \cdots \\
\cdots & \cdots & \ddots & \mathbf{0} \\
\mathbf{0} & \mathbf{0} & \mathbf{0} & \Phi(k_{d,e,m})
\end{matrix}\right],
\end{equation}
where $\Phi$ is the homomorphism from GF($q^L$) into the ring of $L\times L$ matrix over GF($q$) defined in (\ref{eqn:Matrix_Representation_Homomorphism}).
In the same way as to prove Proposition \ref{prop:matrix_representation_scalar_solution}, one can prove that this vector code over GF($q$)$^L$ qualifies as a solution too. We thus obtained the following.

\begin{proposition}
\label{prop_Scalar_Solv_Imply_Vector_Sol}
If a (possibly non-multicast) network is scalar linearly solvable over GF($q^{L_j}$) for all $1 \leq j \leq m$, then it is not necessarily scalar linearly solvable over GF($q^L$), but must be vector linearly solvable over GF($q$)$^L$, where $L = L_1 + L_2 + \cdots + L_m$.
\end{proposition}

As a consequence of the above analysis, the Swirl Network with source dimension $\omega \geq 2^{13} - 2$ has a vector linear solution over GF(2)$^{13}$ but no scalar linear solution over GF($2^{13}$). This satisfies our desired property. Next result can be further obtained.

\begin{proposition}
\label{prop_Swirl_Network_Vector_Solvability}
Let $2^L - 1$ be an arbitrary Mersenne prime no smaller than $2^{13} - 1$. The Swirl Network with source dimension $\omega \geq 2^L - 2$ is vector linearly solvable over GF($q'$)$^{L'}$ for every $q'^{L'} \geq 2^L$, but not scalar linearly solvable over GF($q'$) for any $q'$ with $2^L \leq q' \leq \omega + 2$ and $q' - 1$ being a prime.
\begin{proof}
Corollary \ref{cor:Swirl_Network_Scalar_Solv} characterizes the scalar linear solvability of the considered Swirl network. It remains to show its vector linear solvability. Assume $q'$ is odd. The Swirl Network is scalar linearly solvable over GF($q'^{L'}$) for every $L' \geq 1$, and hence vector linearly solvable over GF($q'$)$^{L'}$ by Proposition \ref{prop:matrix_representation_scalar_solution}. Assume $q' = 2$. The case $L' = 13$ has been discussed in the analysis prior to the present proposition. Consider the case $L' > 13$. Since $2^{L'} - 1$ a Mersenne prime, $L$ is an (odd) prime too. Thus, $L' - 13$ is even and hence $2^{L'-13} - 1$ must be a composite. Consequently, the considered Swirl Network is scalar linearly solvable over GF($2^4$), GF($2^9$) and GF($2^{L'-13}$) respectively. Consequently, it is vector linearly solvable over GF($2$)$L'$ according to Proposition \ref{prop_Scalar_Solv_Imply_Vector_Sol}.
\end{proof}
\end{proposition}

\begin{corollary}
Let $2^L - 1$ be a Mersenne prime no smaller than $2^{13} - 1$, and $\mathcal{N}_1$ represent the Swirl Network with source dimension $\omega \geq 2^L - 2$. The multicast network $\mathcal{N}$ constructed by Algorithm 1 with $n = 2^L$ has a vector linear solution over GF($2$)$^L$, but does not have a scalar linear solution over GF($q'$) for any $q' \leq 2^L$. Moreover, it is vector linearly solvable over GF($q'$)$^{L'}$ for every $q'^{L'}$ larger than $2^L$, but not scalar linearly solvable over GF($q'$) for any $q'$ with $2^L \leq q' \leq \omega + 2$ and $q' - 1$ being a prime.
\begin{proof}
It follows from Theorem \ref{thm:general_construction} together with Proposition \ref{prop_Swirl_Network_Vector_Solvability}.
\end{proof}
\end{corollary}

We have now affirmed the correctness of the conjecture raised in \cite{Fragouli11} by explicit examples that vector linear coding can indeed be superior to scalar one in terms of the required alphabet size in a linear network coding solution. Furthermore, these first exemplifying networks suggest that there are cases where vector linear coding are superior to scalar linear coding in a stronger sense than as conjectured in \cite{Fragouli11}:

\begin{itemize}
\item It is possible for a multicast network vector linearly solvable over GF($q'$)$^{L'}$ for every prime power $q'^{L'} \geq q^L$, but not scalar linearly solvable not only over any GF($q'$) with $q' \leq q^L$, but also over some GF($q'$) with $q' > q^L$, which can be extremely large compared with $q^L$.
\end{itemize}

\subsection{Construction of Infinitely Many Network Instances}
In the previous subsection, the key to proving the Swirl Network to be vector linearly solvable over GF($2$)$^L$ but not scalar linearly solvable over GF($2^L$) is the observation that scalar linear solutions over respective alphabets $\mathrm{GF}(q^{L_1}), \cdots, \mathrm{GF}(q^{L_m})$ do not necessarily imply another scalar linear solution over $\mathrm{GF}(q^{L_1 + \cdots + L_m})$, but they guarantee a vector linear solution over GF($q$)$^{L_1 + \cdots + L_m}$. At this moment, it only brings us a few alphabet sizes $2^L$ with the property that there is a multicast network vector linearly solvable over GF($2$)$^L$ but not scalar linearly solvable over GF($2^L$). In this subsection, we shall identify \emph{infinitely many} alphabet sizes with this property. Towards this goal, we first characterize the vector linear solvability of the network $\mathcal{N}_{\omega, \mathbf{d}}$ described in Section II.B.

\begin{lemma}
\label{lemma:equivalent_condition_VLNC_1}
The network $\mathcal{N}_{\omega,\mathbf{d}}$ with parameters $\omega$ and $\mathbf{d} = (d_1, \cdots, d_\omega)$ has a vector linear solution over GF($q$)$^L$ if and only if there exist invertible matrices $\mathbf{A}_{11}, \cdots, \mathbf{A}_{1d_1}, \cdots, \mathbf{A}_{\omega1}, \cdots, \mathbf{A}_{\omega d_\omega}$ of size $L \times L$ over GF($q$) such that
\begin{align}
\label{eqn:equivalent_condition_1}
 & rank\left(\mathbf{A}_{jk_1} - \mathbf{A}_{jk_2}\right) = L~~~\forall 1 \leq j \leq \omega, 1 \leq k_1 < k_2 \leq d_j, \\
 & rank\left(\mathbf{I} + (-1)^{\omega-1}\mathbf{M}_\omega\mathbf{M}_{\omega-1}\cdots\mathbf{M}_1\right) = L \nonumber \\
\label{eqn:equivalent_condition_2}
 &~~~~~~~~~~\forall~\mathbf{M}_j \in \left\{\mathbf{A}_{j1}, \cdots, \mathbf{A}_{jd_j}\right\}, 1 \leq j \leq \omega.
\end{align}
\end{lemma}

The technical proof of this lemma is provided in Appendix \ref{Appendix:lemma:equivalent_condition_VLNC_1}. It is worthwhile to note that when $L = 1$, the lemma degenerates to a scalar linear solvability characterization of $\mathcal{N}_{\omega,\mathbf{d}}$, which coincides with the one derived in \cite{Sun_15} as a preliminary to further obtain Theorem \ref{thm:Concise_Scalar_Linear_Solvability_Characterization}.

\begin{proposition}
\label{prop:infinitely_many_instances}
Let $l$ be an arbitrary integer larger than 2. Set $\omega \geq 484$ and $\mathbf{d} = \left(\left\lceil\frac{2^{6l+1} - 1}{22}\right\rceil, \left\lceil\frac{2^{6l+1} - 1}{22}\right\rceil, \cdots, \left\lceil\frac{2^{6l+1} - 1}{22}\right\rceil\right)$. Then, $\mathcal{N}_{\omega,\mathbf{d}}$ is vector linearly solvable over GF($2$)$^{6l+1}$ but not scalar linearly solvable over GF($2^{6l+1}$).
\begin{proof}
We first show that the network $\mathcal{N}_{\omega,\mathbf{d}}$ is not scalar linearly solvable over GF($2^{{6l+1}}$). Observe that the smallest $n$ for $2^n \equiv 1 \mod p$ for $p = 3, 5, 7, 11, 13, 17, 19$ is, respectively, $2, 4, 3, 10, 12, 8, 18$. Thus, it can be seen that the smallest positive integer that possibly divides $2^{6l+1} - 1$ is 23, and for every proper divisor $d$ of $2^{6l+1} - 1$, $d \leq \frac{2^{6l+1} - 1}{23} < \frac{2^{6l+1} - 1}{22}$. Consequently,
\begin{align*}
&d\left(\underbrace{\left\lceil\frac{\lceil\frac{2^{6l+1} - 1}{22}\rceil}{d}\right\rceil + \cdots + \left\lceil\frac{\lceil\frac{2^{6l+1} - 1}{22}\rceil}{d}\right\rceil}_{\omega} - \omega + 1 \right) + 2 \\
\geq &d\left(\underbrace{\left\lceil\frac{2^{6l+1} - 1}{22d}\right\rceil + \cdots + \left\lceil\frac{2^{6l+1} - 1}{22d}\right\rceil}_{\omega} - \omega + 1 \right) + 2 \\
> &\omega\left(\frac{2^{6l+1}-1}{22} - d\right) + d + 2 \\
\geq & 484\frac{2^{6l+1} - 1}{22} - 483\frac{2^{6l+1}-1}{23} + 2 = 2^{6l+1} + 1,
\end{align*}
and hence condition (\ref{eqn:equivalence_subgroup_order_perspective}) does not hold. Theorem \ref{thm:Concise_Scalar_Linear_Solvability_Characterization} then affirms that $\mathcal{N}_{\omega,\mathbf{d}}$ is not scalar linearly solvable over GF($2^{6l+1}$).

We next establish a vector linear solution for $\mathcal{N}_{\omega,\mathbf{d}}$ over GF($2$)$^{6l+1}$. Write $d = \left\lceil\frac{2^{6l+1}-1}{22}\right\rceil$, $L = 6l+1$, $L_1 = 9$ and $L_2 = 6l-8$. Let $\mathbf{G}_1$ be the $L_1 \times L_1$ invertible matrix over GF(2) representing a primitive element in GF($2^{L_1}$), and $\mathbf{G}_2$ be the $L_2 \times L_2$ invertible matrix over GF(2) representing a primitive element in GF($2^{L_2}$) according to the homomorphism presented in (\ref{eqn:Matrix_Representation_Homomorphism}) in Section II.A. Then, $rank(\mathbf{G}_1^{j_1} - \mathbf{G}_1^{j_2}) = L_1$ for all $1 \leq j_1 < j_2 \leq 2^{L_1} - 1$, and $rank(\mathbf{G}_2^{j_1} - \mathbf{G}_2^{j_2}) = L_2$ for all $1 \leq j_1 < j_2 \leq 2^{L_2} - 1$.

Write $m_1 = \frac{2^{L_1} - 1}{7}$ and $m_2 = \frac{2^{L_2} - 1}{3}$. Note that both $m_1$ and $m_2$ are integers. For $1 \leq j \leq m_1$, $1 \leq k \leq m_2$, define $\mathbf{B}_{jk}$ to be the $L \times L$ matrix over GF(2) as
\[
\mathbf{B}_{jk} = \left[\begin{matrix} \mathbf{G}_1^{7j} & \mathbf{0} \\ \mathbf{0} & \mathbf{G}_2^{3k} \end{matrix}\right].
\]
Thus, $rank\left(\mathbf{B}_{j_1k_1} - \mathbf{B}_{j_2k_2}\right) = rank\left(\mathbf{G}_1^{7j_1} - \mathbf{G}_1^{7j_2}\right) + rank\left(\mathbf{G}_2^{3k_1} - \mathbf{G}_2^{3k_2}\right) = L$ for all $1 \leq j_1 < j_2 \leq m_1, 1 \leq k_1 < k_2 \leq m_2$. Since
\begin{align*}
m_1m_2 - d &= \frac{2^{9} - 1}{7}\frac{2^{6l-8} - 1}{3} - \left\lceil\frac{2^{6l+1}-1}{22}\right\rceil \\
& > \frac{2^{6l+1}}{21\cdot22} - \frac{2^9 + 2^{6l-8} - 1}{21} - 1 > 0,
\end{align*}
where the last inequality holds as $l$ is assumed larger than $2$, we can set $\mathbf{A}_{n1}, \cdots, \mathbf{A}_{nd}$ to be arbitrary $d$ distinct matrices in $\{\mathbf{B}_{jk}: 1 \leq j \leq m_1, 1 \leq k \leq m_2\}$ for $1 \leq n \leq \omega - 1$, and set $\mathbf{A}_{\omega 1} = \mathbf{A}_0\mathbf{A}_{11}, \mathbf{A}_{\omega 2} = \mathbf{A}_0\mathbf{A}_{12}, \cdots, \mathbf{A}_{\omega d} = \mathbf{A}_0\mathbf{A}_{1d}$, where
\[
\mathbf{A}_0 = \left[\begin{matrix} \mathbf{G}_1 & \mathbf{0} \\ \mathbf{0} & \mathbf{G}_2 \end{matrix}\right].
\]
In this way, $rank\left(\mathbf{A}_{jk_1} - \mathbf{A}_{jk_2}\right) = L$ for all $1 \leq j \leq \omega$ and $1 \leq k_1 < k_2 \leq d$, and
\begin{align*}
\mathbf{I} &\notin \left\{\mathbf{A}_0\mathbf{B}_\omega\cdots\mathbf{B}_2\mathbf{B}_1: \mathbf{B}_n \in \{\mathbf{B}_{jk}, 1 \leq j \leq m_1, 1 \leq k \leq m_2\} \right\} \\
& \supset \left\{\mathbf{M}_\omega\mathbf{M}_{\omega-1}\cdots\mathbf{M}_1: \mathbf{M}_j \in \left\{\mathbf{A}_{j1}, \cdots, \mathbf{A}_{jd}\right\}, 1 \leq j \leq \omega\right\}.
\end{align*}
This means that the designed $\mathbf{A}_{11}, \cdots, \mathbf{A}_{1d}, \cdots, \mathbf{A}_{\omega1}, \cdots, \mathbf{A}_{\omega d}$ satisfy (\ref{eqn:equivalent_condition_1}) and (\ref{eqn:equivalent_condition_2}), so $\mathcal{N}_{\omega, \mathbf{d}}$ is vector linearly solvable over GF($2$)$^L$ according to Lemma \ref{lemma:equivalent_condition_VLNC_1}.
\end{proof}
\end{proposition}

In the proof of $\mathcal{N}_{\omega,\mathbf{d}}$ to be vector linearly solvable over GF($2$)$^{6l+1}$ in Theorem \ref{prop:infinitely_many_instances}, we essentially constructed a scalar linear code over GF($2^9$) and another scalar linear code GF($2^{6l-8}$), none of which qualifies as a solution according to Theorem \ref{thm:Concise_Scalar_Linear_Solvability_Characterization}. Then, we combine their corresponding vector linear codes by direct sum and form a vector linear solution. This provides a new approach to design vector linear codes which scalar codes cannot substitute. By a similar but more elaborate argument, we are able to obtain the following generalization.

\begin{theorem}
\label{thm:arbitrary_p_infinitely_many_instances}
Let $p$ be an arbitrary prime. There exists a positive integer $m$ such that for every $p^{ml+1}$, $l \geq 1$, an instance $\mathcal{N}_{\omega,\mathbf{d}}$ can be found to be vector linearly solvable over GF($p$)$^{ml+1}$ but not scalar linearly solvable over GF($p^{ml+1}$).
\begin{proof}
Please refer to Appendix \ref{Appendx:thm:arbitrary_p_infinitely_many_instances}.
\end{proof}
\end{theorem}

If we let $\mathcal{N}_1$ represent the network $\mathcal{N}_{\omega,\mathbf{d}}$ established in Theorem \ref{thm:arbitrary_p_infinitely_many_instances}, then the multicast network $\mathcal{N}$ constructed by Algorithm 1 with $n = p^{ml+1}$ has a vector linear solution over GF($p$)$^{ml+1}$, but does not have a scalar linear solution over GF($q'$) for any $q' \leq p^{ml+1}$. The conjectured benefit of vector linear codes raised in \cite{Fragouli11} is thus proven in the following more general sense.

\begin{corollary}
For every prime $p$, there are infinitely many alphabet sizes $p^L$ each of which corresponds to a multicast network vector linearly solvable over GF($p$)$^L$ but not scalar linearly solvable over any GF($q'$) with $q' \leq p^L$.
\end{corollary}

\vspace{5pt}

\section{Vector LNC with Smaller Alphabets Better Than Larger Ones}
\begin{figure}[h]
\centering
\scalebox{.88}
{\includegraphics{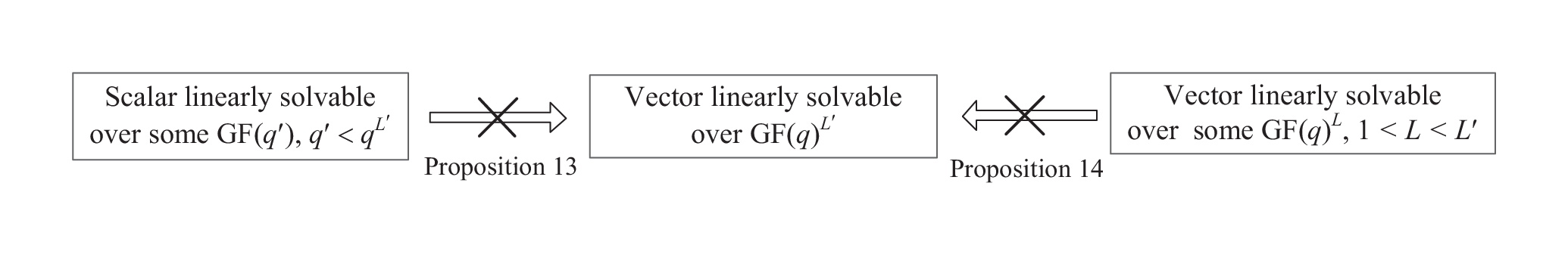}}
\caption{The main results to be established in Section IV.}
\label{Fig:Relation_Diagram_Part2}
\end{figure}

In this section, we shall investigate the vector linear solvability of multicast networks from another direction, in which the main results to be established are outlined in Fig. \ref{Fig:Relation_Diagram_Part2}. %
According to Proposition \ref{prop:matrix_representation_scalar_solution}, we have known that if a multicast network is not vector linearly solvable over GF($q$)$^L$, then it is not scalar linearly solvable over GF($q^L$) either. A natural subsequent question is: when a multicast network is not vector linearly solvable over GF($q$)$^L$, is it scalar linearly unsolvable over GF($q'$) for \emph{any} $q' \leq q^L$? It is tempting to think of a `yes' answer, since vector LNC offers a much larger set for local encoding kernel choices. However, we next prove the \emph{negative} answer to this question, through further study of the Swirl Network.

Consider the Swirl Network $\mathcal{N}_{\omega,\mathbf{d}}$ with $\omega \geq 6$ and $\mathbf{d} = (2, \cdots, 2)$ again. As a consequence of Corollary \ref{cor:Swirl_Network_Scalar_Solv}, it is scalar linearly solvable over both GF(5) and GF(7), no matter how large $\omega$ is selected. We shall next investigate its vector linear solvability.

Our first goal is to check whether the Swirl Network has a vector linear solution over GF(2)$^3$ when $\omega = 6$. Based on Lemma \ref{lemma:equivalent_condition_VLNC_1}, a straightforward way to do so is to exhaustively enumerate all invertible $3\times 3$ matrices over GF(2) for $\mathbf{A}_{11}, \mathbf{A}_{12}, \cdots, \mathbf{A}_{61}, \mathbf{A}_{62}$ to see whether conditions (\ref{eqn:equivalent_condition_1}) and (\ref{eqn:equivalent_condition_2}) hold. However, because there are total $(2^3 - 2^0)(2^3 - 2^1)(2^3-2^2) = 168$ invertible $3 \times 3$ matrices over GF(2), the raw exhaustive enumeration will involve $168^{12}$ combinations, and such computational complexity is too high to realize. In order to reduce the computational complexity in exhaustive enumeration, we are able to further refine the equivalent conditions in Lemma \ref{lemma:equivalent_condition_VLNC_1} for the Swirl Network as follows. Similar refinement can also be conducted for a general $\mathcal{N}_{\omega, \mathbf{d}}$ but we shall not address it in this paper.

\begin{lemma}
\label{lemma:necessary_condition_SwirlNetwork}
The Swirl Network has a vector linear solution over GF($q$)$^L$ if and only if there exist invertible matrices $\mathbf{B}_{1}, \cdots, \mathbf{B}_{\omega}, \mathbf{B}_{\omega+1}$ of size $L \times L$ over GF($q$) such that
\begin{align}
\label{eqn:necessary_condition_1}
&rank\left(\mathbf{I} - \mathbf{B}_{j}\right) = L ~~\forall~1 \leq j \leq \omega, \\
&rank\left(\mathbf{B}_{\omega+1} + \mathbf{M}_{\omega}\mathbf{M}_{\omega-1}\cdots\mathbf{M}_1\right) = L \nonumber \\
\label{eqn:necessary_condition_2}
&~~~~~~~~~~~~~~~~~~\forall~\mathbf{M}_j\in \left\{\mathbf{I}, \mathbf{B}_{j}\right\}, 1 \leq j \leq \omega.
\end{align}
\begin{proof}
Please refer to Appendix \ref{Appendix:lemma:necessary_condition_SwirlNetwork}.
\end{proof}
\end{lemma}

Consider the case that $q = 2$ and $L = 3$. There are total 48 invertible matrices $\mathbf{B}$ over GF(2) of size $3\times 3$ satisfying $rank(\mathbf{I} - \mathbf{B}) = 3$ by computer search. Thus, the number of combinations to search for invertible matrices $\mathbf{B}_1, \cdots, \mathbf{B}_{7}$ subject to (\ref{eqn:necessary_condition_1}) and (\ref{eqn:necessary_condition_2}), when $\omega$ is set to 6, is $48^{7}$, which becomes more manipulable. By a divide-and-conquer method on $\omega$, we first find that there are 2304 sets of invertible matrices $\mathbf{B}_1, \cdots, \mathbf{B}_4$ subject to (\ref{eqn:necessary_condition_1}) and (\ref{eqn:necessary_condition_2}) when $\omega$ is set to 3. Based on this finding, further exhaustive enumeration verifies that no invertible matrices $\mathbf{B}_1, \cdots, \mathbf{B}_{7}$ can be found to make (\ref{eqn:necessary_condition_1}) and (\ref{eqn:necessary_condition_2}) hold. It can also be readily verified that there are not invertible matrices $\mathbf{B}_1, \cdots, \mathbf{B}_{7}$ over GF(2) of size $2\times 2$ to make conditions (\ref{eqn:necessary_condition_1}) and (\ref{eqn:necessary_condition_2}) hold, so the Swirl Network is not vector linearly solvable over GF(2)$^2$ either. In addition, the Swirl Network is not scalar (and thus not vector) linearly solvable over GF(2). Since the Swirl Network with $\omega > 6$ has a vector linear solution over GF($q$)$^L$ only if so is the Swirl Network with $\omega = 6$, we conclude the following.

\begin{proposition}
\label{prop:Swirl_non_vector_solution}
For $\omega \geq 6$, the Swirl Network is scalar linearly solvable over GF(5) and GF(7), but does not have a vector linear solution over GF($2$)$^L$ for any $L \leq 3$.
\end{proposition}

The Swirl Network affirms that even though the choice of local encoding kernels in scalar LNC is more restricted than in vector LNC, scalar LNC can still outperform vector LNC (of dimension larger than 1) in terms of enabling a linear solution using a smaller alphabet. Since every scalar solution can be regarded as a vector solution of dimension 1, this finding suggests that the alphabet size for vector LNC is not always the larger the better for yielding a solution on a multicast network.

Next, we present a more surprising result that over the same base field, a higher dimension of vector LNC is \emph{not} always more likely to enable a linear multicast solution. 

\begin{proposition}
\label{prop:Swirl_non_vector_solution2}
The Swirl Network, which has a scalar linear solution over GF($2^4$) and thus a vector linear solution over GF($2$)$^4$, is \emph{not} vector linearly solvable over GF($2$)$^5$ when source dimension $\omega$ is large enough.
\begin{proof}
According to Corollary \ref{cor:Swirl_Network_Scalar_Solv} and Proposition \ref{prop:matrix_representation_scalar_solution}, it is straightforward to see that the Swirl Network is scalar linearly solvable over GF($2^4$) and then vector linearly solvable over GF($2$)$^4$. In order to show that the Swirl Network is not vector linearly solvable over GF($2$)$^5$, by Lemma \ref{lemma:necessary_condition_SwirlNetwork}, it is equivalent to show the non-existence of invertible matrices $\mathbf{B}_1, \cdots, \mathbf{B}_{\omega + 1}$ of size $5\times 5$ over GF(2) to make conditions (\ref{eqn:necessary_condition_1}) and (\ref{eqn:necessary_condition_2}) hold. However, as $\omega$ is large and there are $(2^5-1)(2^5 - 2)\cdots(2^5-2^4) = 9999360$ invertible matrices of size $5\times 5$ over GF($2$), which form the general linear group $GL_5(2)$, it is hard to directly check this by exhaustive enumeration. By analyzing the group structure of $GL_5(2)$, which is provided in \cite{Atlas_Group}, we shall first greatly reduce the cases to the degree that exhaustively enumeration is manipulable.

Assume that there is a vector linear solution for the Swirl Network over GF($2$)$^5$, and let $\mathbf{B}_1, \cdots, \mathbf{B}_{\omega+1}$ be $5\times 5$ matrices over GF($2$) satisfying conditions (\ref{eqn:necessary_condition_1}) and (\ref{eqn:necessary_condition_2}).

Recall that the conjugacy class of an element $a$ in a group $G$ refers to the set $\{gag^{-1}: g \in G\}$. The elements in a group can be partitioned into conjugacy classes and elements in the same conjugacy class have the same order. %
Since $rank(\mathbf{I} - \mathbf{A}\mathbf{B}\mathbf{A}^{-1}) = rank(\mathbf{A}(\mathbf{I} - \mathbf{B})\mathbf{A}^{-1})$ for every $\mathbf{A} \in GL_5(2)$, $rank(\mathbf{I} - \mathbf{B}) = 5$ for a matrix $\mathbf{B} \in GL_5(2)$ if and only if $rank(\mathbf{I} - \mathbf{B'}) = L$ for every matrix $\mathbf{B'}$ in the conjugacy class of $\mathbf{B}$ in $GL_5(2)$. After examining the representative of every conjugacy class in $GL_5(2)$, as listed in \cite{Atlas_Group}, we found that there are only 8 conjugacy classes in $GL_5(2)$ in which the matrices $\mathbf{B}$ satisfy $rank(\mathbf{I}-\mathbf{B}) = 5$. Two of the 8 conjugacy classes comprise matrices of order 21 in $GL_2(5)$, and the other 6 conjugacy classes comprise matrices of order 31 in $GL_2(5)$. Thus, $\mathbf{B}_1, \cdots, \mathbf{B}_{\omega+1}$ are contained in the union of these $8$ conjugacy classes.

Next, as $\omega$ is assumed large enough,
\[
\{\mathbf{I}, \mathbf{B}_1, \cdots, \mathbf{B}_1^{31}\} \subseteq \{\mathbf{M}_{\omega}\mathbf{M}_{\omega-1}\cdots\mathbf{M}_{1}: \mathbf{M}_j \in \{\mathbf{I}, \mathbf{B}_j\}, 1\leq j \leq \omega \},
\]
and thus $rank(\mathbf{B}_{\omega+1} + \mathbf{B}_{1}^j) = 5$ for all $1 \leq j \leq 31$. Let $\mathbf{B}_1'$ be the representative listed in \cite{Atlas_Group} for the conjugacy class which $\mathbf{B}_1$ belongs to. Then, $\mathbf{B}_1'$ can be written as $\mathbf{A}\mathbf{B}_1\mathbf{A}^{-1}$ for some $\mathbf{A} \in GL_5(2)$. Also write $\mathbf{B}_{\omega+1}' = \mathbf{A}\mathbf{B}_{\omega+1}\mathbf{A}^{-1}$. Then,
\[
rank(\mathbf{B}_{\omega+1}' - \mathbf{B}_{1}'^j) = rank(\mathbf{A}(\mathbf{B}_{\omega+1} - \mathbf{B}_{1}^j)\mathbf{A}^{-1}) = 5 ~~~\forall 1 \leq j \leq 31.
\]

It can be observed that the set $\{\mathbf{B}'_1, \mathbf{B}_1'^2, \cdots, \mathbf{B}_1'^{31}\}$ is identical no matter which conjugacy class of order $31$ matrices $\mathbf{B}'_1$ is in. It can further be checked that $rank(\mathbf{B}_1'^j - \mathbf{B}_1'^k) = 5$ for all $1 \leq j < k \leq 31$. Thus, the $33$ matrices $\mathbf{0}, \mathbf{B}_1, \cdots, \mathbf{B}_1^{31}, \mathbf{B}_{\omega+1}$ form a $5$-dimensional rank-metric code of distance $5$ over GF($2$). However, this contradicts the fact that the number of codewords of such a code is at most $2^5 = 32$ according to the Singleton bound of rank-metric codes.

Consequently, $\mathbf{B}'_1$ can only be the representative for either of the two conjugacy classes of order $21$ matrices. It can be observed that the set $\{\mathbf{B'}_1, \cdots, \mathbf{B'}_1^{21}\}$ is identical for both cases. Then, exhaustive enumeration can be readily conducted on all possible $\mathbf{B'}_{\omega+1} \in GL_5(2)$ with $rank(\mathbf{I} - \mathbf{B'}_{\omega+1}) = 5$ to verify that there does not exist $\mathbf{B'}_{\omega+1}$ such that $rank(\mathbf{B'}_{\omega+1} - \mathbf{B'}_1^j) = 5$ for all $1 \leq j \leq 21$. We can now conclude that there does not exist a vector linear solution over GF($2$)$^5$ when $\omega$ is large enough.
\end{proof}
\end{proposition}

\begin{remark}
It has been proven in \cite{DFZeger07} that the classical network proposed in \cite{Medard03} which is not scalar linearly solvable over any field has a vector linear solution over GF($q$)$^L$ if and only if $L$ is even. However, the discovery in Proposition \ref{prop:Swirl_non_vector_solution2} is more surprising in the sense that a multicast network is considered, which always has a linear solution over a sufficiently larger alphabet. In comparison, a solvable (non-multicast) network is not even vector linearly solvable over any GF($q$)$^L$ \cite{DFZeger05}.
\end{remark}

\vspace{3pt}

\begin{figure}[htbp]
\centering
\scalebox{.73}
{\includegraphics{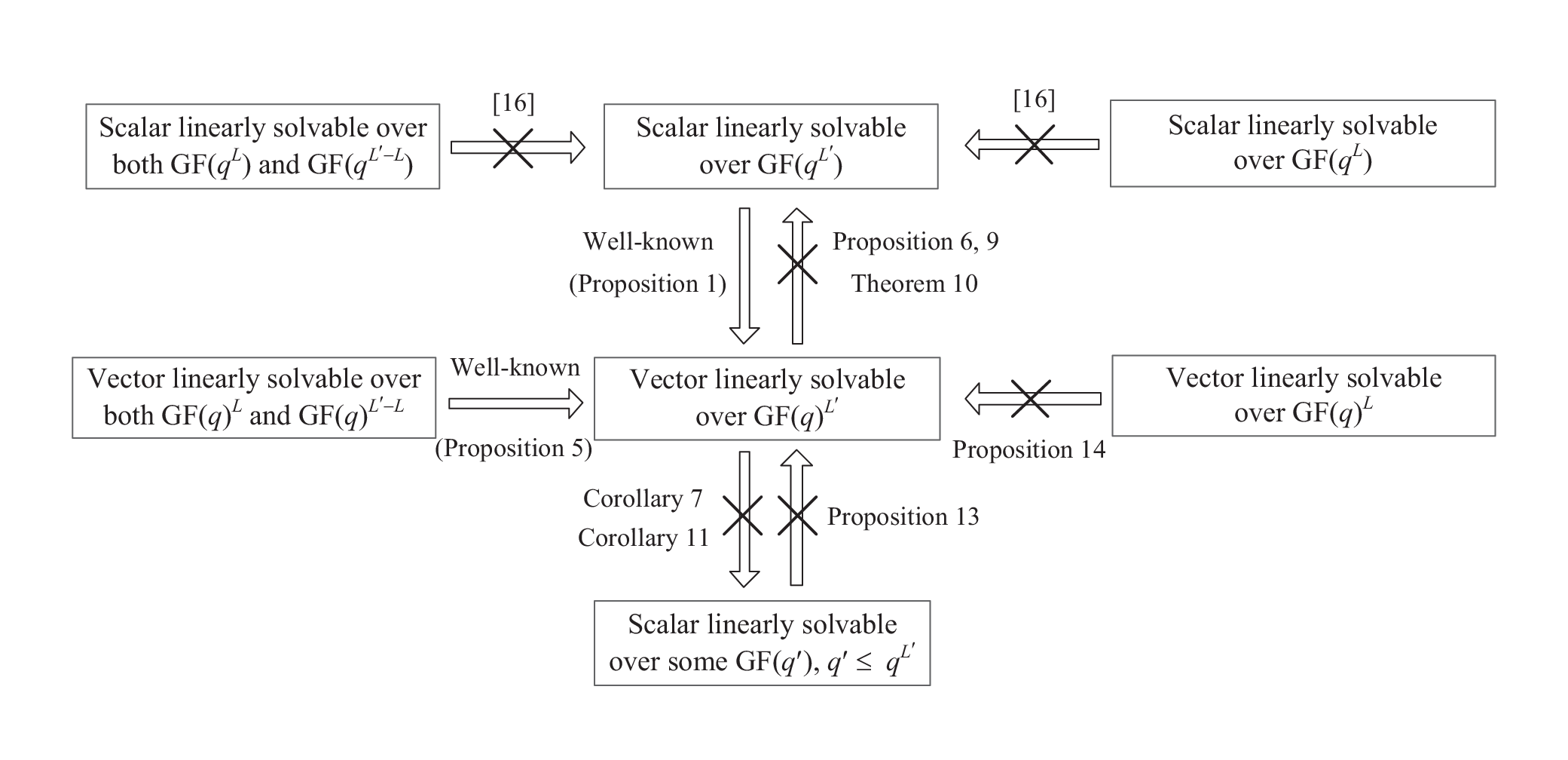}}
\caption{The relationship established between scalar and vector linear solvability on a multicast network. Herein, $1 < hL < L'$.}
\label{Fig:Relationship_between_scalar_vector}
\end{figure}
\section{Summary}
In this paper, several new results are established for vector linear network coding (LNC) on multicast networks. A systematic way is first introduced to construct a multicast network that has a vector linear solution over GF($p$)$^L$, but does not have a scalar linear solution over any GF($q'$) with $q' \leq p^L$, for an \emph{arbitrary} prime $p$ and \emph{infinitely many} alphabet sizes $p^L$. This affirms a conjectured benefit of vector LNC over scalar one in \cite{Fragouli11}. In addition, the technique to construct a vector linear solution is new: a vector linear solution over GF($q$)$^L$ can be constructed by direct sum of different scalar linear codes, which are \emph{not necessarily} scalar linear solutions, over GF($q^{L_1}$), $\cdots$, GF($q^{L_m}$) with $L_1 + \cdots L_m = L$. This is demonstrated to be useful and cannot be substituted by scalar LNC because a multicast network which has scalar linear solutions over GF($q^{L_1}$), $\cdots$, GF($q^{L_m}$) is not even scalar linearly solvable over GF($q^{L_1 + \cdots + L_m}$). In the second part of the paper, explicit multicast networks are presented and proven to have the special property that they do not have a vector linear solution of dimension $L$ over GF($2$) but have scalar linear solutions over GF($q'$) for some $q' < 2^L$, where $q'$ can be odd or even. This discovery unveils a surprising result for vector LNC on multicast networks: the existence of a vector linear solution over GF($2$)$^L$ does not imply the existence of a vector linear solution over GF($2$)$^{L'}$ with $L' > L$. Fig. \ref{Fig:Relationship_between_scalar_vector} summarizes the relationship between scalar and vector LNC on a multicast network established so far in the literature.

\appendix
\subsection{Proof of Lemma \ref{lemma:equivalent_condition_VLNC_1}}
\label{Appendix:lemma:equivalent_condition_VLNC_1}
Denote by $e_{j1}, \cdots, e_{jd_j}$ the $d_j$ outgoing edges of layer-3 node $v_j$ for each $1 \leq j \leq \omega$. Consider a vector linear code over GF($q$)$^L$. For $1 \leq j \leq \omega$, denote by $\mathbf{U}_{j1}, \cdots, \mathbf{U}_{jd_j}$ the local encoding kernels for adjacent pairs $(d_{j}, e)$ with $e \in In(v_j)$ and $e \in In(v_{j-1})$, respectively, where $v_0$ stands for $v_\omega$. Note that by left multiplying $\mathbf{U}_{j1}$ to the local encoding kernels for downstream adjacent pairs $(e, e_{j1}), \cdots, (e, e_{jd_j})$, and resetting $\mathbf{U}_{j1}$ to be the $L \times L$ identity matrix $\mathbf{I}$, the global encoding kernels for edges $e_{j1}, \cdots, e_{jd_j}$ remain unchanged. Hence, without loss of generality, for all $1 \leq j \leq \omega$ and $1 \leq k \leq d_j$, we can assume $\mathbf{U}_{jk} = \mathbf{I}$ and let $\mathbf{K}_{jk}, \mathbf{K}_{jk}'$ denote the local encoding kernels for $(e, e_{jk})$, $e \in In(v_j)$. Then, the juxtaposition of global encoding kernels for edges $e_{jk}$ is equal to
\begin{equation*}
[\mathbf{F}_{e_{j1}} \cdots \mathbf{F}_{e_{jd_j}}]_{1\leq j \leq \omega}
 = \left[ \begin{smallmatrix}
    \mathbf{K}_{11} & & \mathbf{K}_{1d_1} & & \mathbf{0} & & \mathbf{0} & \mathbf{K}_{\omega d_1} & & \mathbf{K}_{\omega d_\omega} \\
    \mathbf{K}_{11}' & & \mathbf{K}_{1d_1}' & & \vdots & & \vdots & \mathbf{0} & & \mathbf{0} \\
    \mathbf{0} & \cdots & \mathbf{0} & \ddots & \mathbf{0} & \cdots & \mathbf{0} & \vdots & \cdots & \vdots \\
    \vdots & & \vdots &  & \mathbf{K}_{(\omega-1)1} & & \mathbf{K}_{(\omega-1)d_{\omega-1}} & \mathbf{0} & &  \mathbf{0} \\
    \mathbf{0} & & \mathbf{0} &  & \mathbf{K}_{(\omega-1)1}' & &  \mathbf{K}_{(\omega-1)d_{\omega-1}}' & \mathbf{K}_{\omega1}' & &  \mathbf{K}_{\omega d_\omega}' \\
  \end{smallmatrix}
\right]
\end{equation*}
Since there is a receiver connected from every set $N$ of $\omega$ grey nodes with $maxflow(N)$ and each $e_{jk}$ is the unique incoming edge to a grey node, the vector linear code is a solution if and only if for every set $E$ of $\omega$ edges in $\{e_{jk}: 1 \leq j \leq \omega, 1 \leq k \leq d_j\}$ with $maxflow(E) = \omega$, where $maxflow(E)$ means the number edge-disjoint paths starting from the source and ending at edges in $E$, the matrix $[\mathbf{F}_e]_{e \in E}$ is of full rank $\omega L$.

To prove the necessity part of the lemma, assume that the considered code is a vector linear solution.
First observe that $maxflow(E) = \omega$ when $E = \{e_{1k_1}, \cdots, e_{(\omega-2)k_{\omega-2}}, e_{(\omega-1)1}, e_{(\omega-1)2}\}$, where $1 \leq k_j \leq d_j$. Then the matrix $[\mathbf{F}_e]_{e\in E} = \left[ \begin{smallmatrix} \mathbf{K}_{1k_1} & & & \mathbf{0} & \mathbf{0} \\
\mathbf{K}_{1k_1}' & & \cdots & \cdots & \cdots \\
\mathbf{0} & \cdots & \mathbf{K}_{(\omega-2)k_{\omega-2}} & \mathbf{0} & \mathbf{0} \\
\cdots & & \mathbf{K}'_{(\omega-2)k_{\omega-2}} & \mathbf{K}_{(\omega-1)1} & \mathbf{K}_{(\omega-1)2}  \\
\mathbf{0} &  & \mathbf{0} &  \mathbf{K}_{(\omega-1)1}' & \mathbf{K}_{(\omega-1)2}'  \end{smallmatrix} \right]$
has full rank $\omega L$. Because
\begin{align*}
0 &\neq \det([\mathbf{F}_e]_{e\in E}) \\
&=
\det\left(\left[\begin{smallmatrix} \mathbf{K}_{1k_1} & & \mathbf{0} & \mathbf{0} \\
\mathbf{K}_{1k_1}' & & \cdots & \cdots \\
\cdots & \cdots & \mathbf{K}_{(\omega-3)k_{\omega-3}} & \mathbf{0}  \\
\mathbf{0} & & \mathbf{K}_{(\omega-3)k_{\omega-3}}' & \mathbf{K}_{(\omega-2)k_{\omega-2}}  \end{smallmatrix} \right]\right)
\det\left(\left[\begin{smallmatrix} \mathbf{K}_{(\omega-1)1} & \mathbf{K}_{(\omega-1)2} \\ \mathbf{K}_{(\omega-1)1}' & \mathbf{K}_{(\omega-1)2}' \end{smallmatrix}\right]\right) \\
&= \det(\mathbf{K}_{1k_1})\cdots\det(\mathbf{K}_{(\omega-2)k_{\omega-2}})\det\left(\left[\begin{smallmatrix} \mathbf{K}_{(\omega-1)1} & \mathbf{K}_{(\omega-1)2} \\ \mathbf{K}_{(\omega-1)1}' & \mathbf{K}_{(\omega-1)2}' \end{smallmatrix}\right]\right),
\end{align*}
the local encoding kernels $\mathbf{K}_{1k_1}, \cdots, \mathbf{K}_{(\omega-2)k_{\omega-2}}$ are invertible matrices. By similar arguments on the set $\{e_{1k_1}, \cdots, e_{(\omega-2)k_{\omega-2}}, e_{\omega1}, e_{\omega2}\}$, $\{e_{11}, e_{12}, e_{3k_3}, \cdots, e_{\omega k_\omega}\}$, and $\{e_{21}, e_{22}, e_{3k_3}, \cdots, e_{\omega k_\omega}\}$, where $1 \leq k_j \leq d_j$, we can deduce that all local encoding kernels $\mathbf{K}_{jk}$ and $\mathbf{K}_{jk}'$ for $1 \leq j \leq \omega$, $1 \leq k \leq d_j$, are invertible matrices.

Write $\mathbf{A}_{jk} = \mathbf{K}_{jk}'\mathbf{K}_{jk}^{-1}$ for $1 \leq j \leq \omega-1$, $1 \leq k \leq d_j$, and $\mathbf{A}_{\omega k} = \mathbf{K}_{\omega k}\mathbf{K}_{\omega k}'^{-1}$ for $1 \leq k \leq d_\omega$. We need show these invertible matrices satisfy conditions (\ref{eqn:equivalent_condition_1}) and (\ref{eqn:equivalent_condition_2}).
Define another vector linear code of dimension $L$ over GF($q$) prescribed by the following global encoding kernels
\begin{equation}
\label{eqn:Swirl_Network_GEK}
[\mathbf{F}_{e_{j1}}' \cdots \mathbf{F}_{e_{jd_j}}']_{1\leq j \leq \omega}
= \left[ \begin{smallmatrix}
    \mathbf{I} & & \mathbf{I} & & \mathbf{0} & & \mathbf{0} & \mathbf{A}_{\omega1} & & \mathbf{A}_{\omega d_\omega} \\
    \mathbf{A}_{11} & & \mathbf{A}_{1 d_1} & & \vdots & & \vdots & \mathbf{0} & & \mathbf{0} \\
    \mathbf{0} & \cdots & \mathbf{0} & \ddots & \mathbf{0} & \cdots & \mathbf{0} & \vdots & \cdots \vdots \\
    \vdots & & \vdots &  & \mathbf{I} & & \mathbf{I} & \mathbf{0} & & \mathbf{0} \\
    \mathbf{0} & & \mathbf{0} &  & \mathbf{A}_{(\omega-1)1} & & \mathbf{A}_{(\omega-1) d_{\omega-1}} & \mathbf{I} & & \mathbf{I} \\
  \end{smallmatrix}
\right]
\end{equation}
Since $[\mathbf{F}_{e_{j1}}' \cdots \mathbf{F}_{e_{jd_j}}']_{1\leq j \leq \omega} = [\mathbf{F}_{e_{j1}} \cdots \mathbf{F}_{e_{jd_j}}]_{1\leq j \leq \omega}\mathbf{Diag}(\ast)$, where $\mathbf{Diag}(\ast)$ stands for the square block diagonal matrix with diagonal blocks equal to $\mathbf{K}_{11}^{-1}, \cdots, \mathbf{K}_{1d_1}^{-1}$, $\cdots$, $\mathbf{K}_{(\omega-1)1}^{-1}, \mathbf{K}_{(\omega-1)d_{\omega-1}}^{-1}, \mathbf{K}_{\omega1}'^{-1}, \mathbf{K}_{\omega d_\omega}'^{-1}$, we have, for any set $E$ of $\omega$ edges with $maxflow(E) = \omega$,
\[rank\left([\mathbf{F}_{e}]_{e\in E}\right) = rank\left([\mathbf{F}_{e}']_{e\in E}\right).\]

Consider $E = \{e_{1k_1}, e_{1k_2}, e_{21}, \cdots, e_{(\omega-1)1}\}$ subject to $maxflow(E) = \omega$, where $1 \leq k_1 < k_2 \leq d_1$. Then,
\begin{align*}
0 &\neq \det\left([\mathbf{F}_e']_{e\in E}\right) = \det\left(\left[
\begin{smallmatrix}
\mathbf{I} & \mathbf{I} & \mathbf{0} & & \mathbf{0} \\
\mathbf{A}_{1k_1} & \mathbf{A}_{1k_2} & \mathbf{I} & & \cdots \\
\mathbf{0} & \mathbf{0} & \mathbf{A}_{21} & \cdots & \mathbf{0} \\
\cdots & \cdots & \cdots & & \mathbf{I} \\
\mathbf{0} & \mathbf{0} & \mathbf{0} & & \mathbf{A}_{(\omega-1)1}
\end{smallmatrix}
\right]\right) \\
&= \det\left(\left[\begin{smallmatrix} \mathbf{I} & \mathbf{I} \\ \mathbf{A}_{1k_1} & \mathbf{A}_{1k_2} \end{smallmatrix}\right]\right)
\det\left(\left[\begin{smallmatrix} \mathbf{A}_{21} & & \mathbf{0} & \mathbf{0} \\ \mathbf{0} & \cdots & \cdots \\ \cdots & & \mathbf{A}_{(\omega-2)1} & \mathbf{I} \\ \mathbf{0} & & \mathbf{0} & \mathbf{A}_{(\omega-1)1} \end{smallmatrix}\right]\right).
\end{align*}
This implies that $0 \neq \det\left(\left[\begin{smallmatrix} \mathbf{I} & \mathbf{I} \\ \mathbf{A}_{1k_1} & \mathbf{A}_{1k_2} \end{smallmatrix}\right]\right) = \det\left(\left[\begin{smallmatrix} \mathbf{I} & \mathbf{0} \\ \mathbf{A}_{1k_1} & \mathbf{A}_{1k_2}-\mathbf{A}_{1k_1} \end{smallmatrix}\right]\right)$, \emph{i.e.}, $rank(\mathbf{A}_{1k_1}-\mathbf{A}_{1k_2}) = L$. By similar arguments on $E = \{e_{jk_1}, e_{jk_2}, e_{(j+1)1}, \cdots, e_{(j+\omega-1)1}\}$, where $1 \leq j \leq \omega$, $1 \leq k_1 < k_2 \leq d_j$, and $e_{l1}$ refers to $e_{(l-\omega)1}$ whenever $l > \omega$, we can verify that condition (\ref{eqn:equivalent_condition_1}) holds for the considered $\mathbf{A}_{jk}$ .

Consider $E = \{e_{1k_1}, e_{2k_2}, \cdots, e_{\omega k_\omega}\}$ subject to $maxflow(E) = \omega$, where $1 \leq k_j \leq d_j$. Then, $[\mathbf{F}_e']_{e\in E} = \left[
\begin{smallmatrix}
\mathbf{I} & \mathbf{0} & & \mathbf{0} & \mathbf{A}_{\omega k_\omega}\\
\mathbf{A}_{1 k_1} & \mathbf{I} & & \cdots & \mathbf{0} \\
\mathbf{0} & \mathbf{A}_{2k_2} & \cdots & \mathbf{0} & \cdots \\
\cdots & \cdots & & \mathbf{I} & \mathbf{0} \\
\mathbf{0} & \mathbf{0} & & \mathbf{A}_{(\omega-1)k_{\omega-1}} & \mathbf{I}
\end{smallmatrix}
\right]$ has full rank $\omega L$.
Set $\mathbf{M}_\omega = \left[\begin{smallmatrix}
\mathbf{I} & & \mathbf{0} & -\mathbf{A}_{\omega k_\omega}\\
\mathbf{0} & & \cdots & \mathbf{0} \\
\cdots & \cdots & \mathbf{0} & \cdots \\
\cdots & & \mathbf{I} & \mathbf{0} \\
\mathbf{0} & & \mathbf{0} & \mathbf{I}
\end{smallmatrix}\right],
\mathbf{M}_{\omega-1} = \left[\begin{smallmatrix}
\mathbf{I} & \mathbf{0} & & \mathbf{A}_{\omega k_\omega}\mathbf{A}_{(\omega-1)k_{\omega-1}} & \mathbf{0}\\
\mathbf{0} & \mathbf{I} & & \cdots & \mathbf{0} \\
\cdots & \mathbf{0} & \cdots & \mathbf{0} & \cdots \\
\cdots & \cdots & & \mathbf{I} & \mathbf{0} \\
\mathbf{0} & \mathbf{0} & & \mathbf{0} & \mathbf{I}
\end{smallmatrix}\right],\cdots$,
$\mathbf{M}_{2} = \left[\begin{smallmatrix}
\mathbf{I} & (-1)^{\omega-1}\mathbf{A}_{\omega k_\omega}\mathbf{A}_{(\omega-1)k_{\omega-1}}\cdots\mathbf{A}_{2k_2} & & \mathbf{0} & \mathbf{0}\\
\mathbf{0} & \mathbf{I} & & \cdots & \cdots \\
\cdots & \mathbf{0} & \cdots & \mathbf{0} & \cdots \\
\cdots & \cdots & & \mathbf{I} & \mathbf{0} \\
\mathbf{0} & \mathbf{0} & & \mathbf{0} & \mathbf{I}
\end{smallmatrix}\right]$. %
Obviously $\det(\mathbf{M}_2) = \cdots =  \det(\mathbf{M}_{\omega-1})= \det(\mathbf{M}_\omega) = 1$. A careful calculation yields
\[
\mathbf{M}_2\cdots \mathbf{M}_\omega[\mathbf{F}_e']_{e\in E} = \left[
\begin{smallmatrix}
\mathbf{I} + (-1)^{\omega-1}\mathbf{A}_{\omega k_\omega}\cdots\mathbf{A}_{1 k_1}& \mathbf{0} & & \mathbf{0} & \mathbf{0}\\
\mathbf{A}_{1 k_1} & \mathbf{I} & & \cdots & \mathbf{0} \\
\mathbf{0} & \mathbf{A}_{2 k_2} & \cdots & \mathbf{0} & \cdots \\
\cdots & \cdots & & \mathbf{I} & \mathbf{0} \\
\mathbf{0} & \mathbf{0} & & \mathbf{A}_{(\omega-1)k_{\omega-1}} & \mathbf{I}
\end{smallmatrix}
\right].
\]
This implies $\det(\mathbf{I} + (-1)^{\omega-1}\mathbf{A}_{\omega k_\omega}\cdots\mathbf{A}_{1 k_1}) = \det([\mathbf{F}_e']_{e\in E}) \neq 0$, and hence $rank(\mathbf{I} + (-1)^{\omega-1}\mathbf{A}_{\omega k_\omega}\cdots\mathbf{A}_{1 k_1}) = L$. Condition (\ref{eqn:equivalent_condition_2}) thus holds for the considered $\mathbf{A}_{jk}$. The necessity part of the lemma is proved.

For the sufficiency part, let $\mathbf{A}_{jk}$, $1 \leq j \leq \omega$, $1 \leq k \leq d_j$, be invertible matrices of size $L \times L$ over GF($q$) subject to conditions (\ref{eqn:equivalent_condition_1}) and (\ref{eqn:equivalent_condition_2}). Assume that the considered vector linear code has local encoding kernels $\mathbf{K}_{jk} = \mathbf{I}$ and $\mathbf{K}_{jk}' = \mathbf{A}_{jk}$ when $1 \leq j \leq \omega-1$, and $\mathbf{K}_{jk}' = \mathbf{I}$, $\mathbf{K}_{jk} = \mathbf{A}_{jk}$ when $j = \omega$, where $1 \leq k \leq d_j$. In this way, the juxtaposition $[\mathbf{F}_{e_{j1}} \cdots \mathbf{F}_{e_{jd_j}}]_{1\leq j \leq \omega}$ of global encoding kernels for edges $\{e_{jk}: 1 \leq j \leq \omega, 1 \leq k \leq d_j\}$ is identical to (\ref{eqn:Swirl_Network_GEK}). It can then be shown, by similar classified discussion following (\ref{eqn:Swirl_Network_GEK}), that for every set $E$ of $\omega$ edges in $\{e_{jk}: 1 \leq j \leq \omega, 1 \leq k \leq d_j\}$ with $maxflow(E) = \omega$, the $\omega L \times \omega L$ matrix $[\mathbf{F}_e]_{e \in E}$ is of full rank $\omega L$. The considered code thus qualifies as a solution for $\mathcal{N}_{\omega,\mathbf{d}}$.

\subsection{Proof of Theorem \ref{thm:arbitrary_p_infinitely_many_instances}}
\label{Appendx:thm:arbitrary_p_infinitely_many_instances}
The case $p = 2$ has been considered in Proposition \ref{prop:infinitely_many_instances}.

Assume that $p$ is odd and $l$ is an arbitrary positive integer. Write $a = p^2 + p + 1$ and $b = 2(p - 1)$. Note that $a$ divides $p^{3l} - 1$ but does not divide $p^{3l+1}-1$, and $b$ divides $p^{2l} - 1$ but does not divide $p^{2l+1} - 1$. Label all odd primes smaller than $ab$ as $p_1, \cdots, p_n$ for some $n$. For each $1 \leq j \leq n$, denote by $q_j$ the smallest power of $p_j$ that does not divide $p - 1$, and by $m_j$ the smallest positive integer subject to $p^{m_j} \equiv 1 \mod q_j$. Define $m$ to be the least common multiplier of $12$ and $m_1, \cdots, m_n$. In this manner, each of $a$, $b$, $q_1, \cdots, q_n$ divides $p^{ml}-1$, but none of them divides $p^{ml+1} - 1$. Moreover, as $\frac{q_1 \cdots q_n}{p_1 \cdots p_n} < p - 1$, the largest divisor of $p^{ml+1} - 1$ that is smaller than $ab$ is $b/2 = p - 1$.

Write $L = ml + 1$, which is no smaller than 13. Denote $\frac{(p^{9} - 1)(p^{L - 9}-1)}{ab}$ by $d_0$, which is always an integer as $a$ divides $p^{9} - 1$ and $b$ divides $p^{L - 9} - 1$. Consider the network $\mathcal{N}_{\omega, \mathbf{d}}$ with $\omega$ sufficiently large and $\mathbf{d} = (\underbrace{d_0, \cdots, d_0}_{\omega-1}, (a-1)(b-1)d_0)$. It suffices to show that it is not scalar linearly solvable over GF($p^L$) but vector linearly solvable over GF($p$)$^L$.

Let $d$ be a proper divisor of $p^L - 1$. In the case $d < d_0$, it is obvious to observe that condition (\ref{eqn:equivalence_subgroup_order_perspective}) does not hold for the $\omega$-tuple $\mathbf{d}$ when $\omega$ is sufficiently large. Consider the case $d \geq d_0$. As $L \geq 13$ and $ab = 2(p^3 - 1)$, it can be readily checked that $\frac{(p^{9} - 1)(p^{L - 9}-1)}{ab} > \frac{p^{L}-1}{ab+1}$, so $d > \frac{p^{L}-1}{ab+1}$ and thus can be written as $(p^L-1)/d'$ for some divisor $d'$ of $p^L - 1$ that is smaller than $ab$. As argued previously, $d' \leq b/2 = p - 1$, so
\begin{align*}
&(a-1)(b-1)d_0 - (d'-1)\frac{p^L-1}{d'} \\
\geq &(a - 1)(b - 1)\frac{(p^9-1)(p^{L-9}-1)}{ab} - \frac{b - 2}{b}(p^L - 1) \\
= &\frac{a-b+1}{ab}p^L - \frac{ab - a - b + 1}{ab}(p^9 + p^{L-9} - 1) + \frac{b-2}{b} \\
> &\frac{2}{ab}p^L - p^9 - p^{L-9} > 0.
\end{align*}
This implies $\left\lceil\frac{(a-1)(b-1)d_0}{d}\right\rceil = d'$. Consequently,
\[
d\left(\underbrace{\left\lceil\frac{d_0}{d}\right\rceil + \cdots + \left\lceil\frac{d_0}{d}\right\rceil}_{\omega-1} + \left\lceil\frac{(a-1)(b-1)d_0}{d}\right\rceil - \omega + 1 \right) + 2
= \frac{p^L-1}{d'}d' + 2 > p^L,
\]
so condition (\ref{eqn:equivalence_subgroup_order_perspective}) does not hold for the case $d \geq d_0$ either. Theorem \ref{thm:Concise_Scalar_Linear_Solvability_Characterization} then asserts that the considered network $\mathcal{N}_{\omega,\mathbf{d}}$ is not scalar linearly solvable over GF($p^L$).

We next establish a vector linear solution for $\mathcal{N}_{\omega,\mathbf{d}}$ over GF($p$)$^{L}$. Let $\mathbf{G}_1$ be the $9 \times 9$ invertible matrix over GF($p$) representing a primitive element in GF($p^{9}$), and $\mathbf{G}_2$ be the $(L-9)\times (L-9)$ invertible matrix over GF($p$) representing a primitive element in GF($p^{L-9}$) according to the homomorphism presented in (\ref{eqn:Matrix_Representation_Homomorphism}) in Section II.A. Then, $rank(\mathbf{G}_1^{j_1} - \mathbf{G}_1^{j_2}) = 9$ for all $1 \leq j_1 < j_2 \leq p^{9} - 1$, and $rank(\mathbf{G}_2^{j_1} - \mathbf{G}_2^{j_2}) = L-9$ for all $1 \leq j_1 < j_2 \leq p^{L-9} - 1$. For $1 \leq j \leq \frac{p^9-1}{a}$, $1 \leq k \leq \frac{p^{L-9}-1}{b}$, define $\mathbf{B}_{jk}$ to be the $L \times L$ matrix $G$ over GF($q$) as %
$\mathbf{B}_{jk} = \left[\begin{matrix} \mathbf{G}_1^{aj} & \mathbf{0} \\ \mathbf{0} & \mathbf{G}_2^{bk} \end{matrix}\right]$. %
Then, $rank(\mathbf{B}_{j_1k_1} - \mathbf{B}_{j_2k_2}) = rank(\mathbf{G}_1^{aj_1} - \mathbf{G}_1^{aj_2}) + rank(\mathbf{G}_2^{bk_1} - \mathbf{G}_2^{bk_2}) = L$ for all $1 \leq j_1 < j_2 \leq \frac{p^9-1}{a}$, $1 \leq k_1 < k_2 \leq \frac{p^{L-9}-1}{b}$.
Set $\mathbf{A}_{n1}, \cdots, \mathbf{A}_{nd_0}$ to be the $d_0$ distinct matrices in $\{\mathbf{B}_{jk}: 1 \leq j \leq \frac{p^9-1}{a}, 1 \leq k \leq \frac{p^{L-9}-1}{b}\}$ for all $1 \leq n \leq \omega - 1$, and set $\mathbf{A}_{\omega 1}, \cdots, \mathbf{A}_{\omega((a-1)(b-1)d_0)}$ to be the $(a-1)(b-1)d_0$ distinct matrices in $\left\{(-1)^{\omega}\left[ %
\begin{smallmatrix}
\mathbf{G}_1^{a'} & \mathbf{0} \\
\mathbf{0} & \mathbf{G}_2^{b'} \\
\end{smallmatrix}
\right]\mathbf{B}_{jk}: 1 \leq a' < a, 1 \leq b' < b, 1 \leq j \leq \frac{p^{9}-1}{a}, 1 \leq k \leq \frac{p^{L-9}-1}{b}\right\}$.
Condition (\ref{eqn:equivalent_condition_1}) naturally holds for thus defined $\mathbf{A}_{jk}$. Moreover, as $\mathbf{G}_1^{aj+a'} \neq \mathbf{I}$ and $\mathbf{G}_2^{bj+b'} \neq \mathbf{I}$, $rank(\mathbf{I} - \mathbf{G}_1^{aj+a'}) = 9$ and $rank(\mathbf{I} - \mathbf{G}_2^{bj+b'}) = L-9$ for all $j \geq 0$, $1 \leq a' < a$ and $1 \leq b < b'$. Consequently, condition (\ref{eqn:equivalent_condition_2}) holds for the defined $\mathbf{A}_{jk}$ too. According to Lemma \ref{lemma:equivalent_condition_VLNC_1}, $\mathcal{N}_{\omega, \mathbf{d}}$ is vector linearly solvable over GF($p$)$^L$.

\subsection{Proof of Lemma \ref{lemma:necessary_condition_SwirlNetwork}}
\label{Appendix:lemma:necessary_condition_SwirlNetwork}
Given invertible matrices $\mathbf{A}_{11}, \mathbf{A}_{12}, \cdots, \mathbf{A}_{\omega1}, \mathbf{A}_{\omega2}$ over GF($q$) of size $L \times L$, define invertible $L \times L$ matrices $\mathbf{B}_j$, $1\leq j \leq \omega+1$ in the following way:
\begin{equation}\label{eqn:matrix_transform}
\begin{matrix}
\mathbf{B}_1 = \left(\mathbf{A}_{11}^{-1}\right)\mathbf{A}_{12}, \\
\mathbf{B}_2 = \left(\mathbf{A}_{11}^{-1}\mathbf{A}_{21}^{-1}\right)\mathbf{A}_{22}\left(\mathbf{A}_{11}\right), \\
\vdots \\
\mathbf{B}_\omega = \left(\mathbf{A}_{11}^{-1}\cdots\mathbf{A}_{\omega1}^{-1}\right)\mathbf{A}_{\omega2}\left(\mathbf{A}_{(\omega-1)1}\cdots\mathbf{A}_{11}\right), \\
\mathbf{B}_{\omega+1} = (-1)^{\omega-1}\mathbf{A}_{11}^{-1}\cdots\mathbf{A}_{\omega1}^{-1}.
\end{matrix}
\end{equation}
Conversely, given invertible $L \times L$ matrices $\mathbf{B}_j$, $1\leq j \leq \omega+1$, define $\mathbf{A}_{11}, \mathbf{A}_{12}, \cdots, \mathbf{A}_{\omega1}, \mathbf{A}_{\omega2}$ to be arbitrary matrices satisfying (\ref{eqn:matrix_transform}). Such a selection always exists because we can set $\mathbf{A}_{11}, \cdots, \mathbf{A}_{(\omega-1)1} = \mathbf{I}$, $\mathbf{A}_{\omega1} = (-1)^{\omega-1}\mathbf{B}_{\omega+1}^{-1}$, and $\mathbf{A}_{j2} = (\mathbf{A}_{j1}\cdots\mathbf{A}_{11})\mathbf{B}_{j}(\mathbf{A}_{11}^{-1}\cdots\mathbf{A}_{(j-1)1}^{-1})$ for $1 \leq j \leq \omega$.

It can be readily checked that the two sets of matrices
\[
\left\{\mathbf{M}_\omega\mathbf{M}_{\omega-1}\cdots\mathbf{M}_{1}: \mathbf{M}_j \in \left\{\mathbf{A}_{j1}, \mathbf{A}_{j2}\right\}, 1\leq j \leq \omega \right\}
\]
and
\[
\left\{\mathbf{A}_{\omega1}\cdots\mathbf{A}_{11}\mathbf{M}_\omega\mathbf{M}_{\omega-1}\cdots\mathbf{M}_{1}: \mathbf{M}_j \in \left\{\mathbf{I}, \mathbf{B}_{j}\right\}, 1\leq j \leq \omega\right\}
\]
are identical. Then,
\begin{align*}
&~~~\left\{\mathbf{I} + (-1)^{\omega-1}\mathbf{M}_\omega\cdots\mathbf{M}_1: \mathbf{M}_j \in \left\{\mathbf{A}_{j1}, \mathbf{A}_{j2}\right\}, 1 \leq j \leq \omega \right\} \\
&= \left\{\mathbf{B}_{\omega+1}^{-1}\left(\mathbf{B}_{\omega+1} + \mathbf{M}_\omega\cdots\mathbf{M}_1\right): \mathbf{M}_j \in \left\{\mathbf{I}, \mathbf{B}_{j}\right\}, 1 \leq j \leq \omega\right\}
\end{align*}
Hence, condition (\ref{eqn:equivalent_condition_2}) holds for $\mathbf{A}_{11}, \mathbf{A}_{12}, \cdots, \mathbf{A}_{\omega1}, \mathbf{A}_{\omega2}$ if and only if condition (\ref{eqn:necessary_condition_2}) holds for $\mathbf{B}_1, \cdots, \mathbf{B}_\omega, \mathbf{B}_{\omega+1}$.
Moreover, because for every $1 \leq j \leq \omega$,
\begin{align*}
&~~~rank(\mathbf{I} - \mathbf{B}_j) \\
&= rank\left(\mathbf{A}_{11}^{-1}\cdots\mathbf{A}_{j1}^{-1}\left(\mathbf{A}_{j1}\cdots\mathbf{A}_{11} - \mathbf{A}_{j2}\mathbf{A}_{(j-1)1}\cdots\mathbf{A}_{11}\right)\right) \\
& = rank\left(\mathbf{A}_{11}^{-1}\cdots\mathbf{A}_{j1}^{-1}\left(\mathbf{A}_{j1} - \mathbf{A}_{j2}\right)\mathbf{A}_{(j-1)1}\cdots\mathbf{A}_{11}\right) \\
& = rank\left(\mathbf{A}_{j1} - \mathbf{A}_{j2}\right),
\end{align*}
condition (\ref{eqn:necessary_condition_1}) holds for $\mathbf{B}_1, \cdots, \mathbf{B}_\omega, \mathbf{B}_{\omega+1}$ if and only if condition (\ref{eqn:equivalent_condition_1}) holds for $\mathbf{A}_{11}, \mathbf{A}_{12}, \cdots, \mathbf{A}_{\omega1}, \mathbf{A}_{\omega2}$.

\section*{Acknowledgment}
The authors would appreciate the helpful suggestions by the editor as well as anonymous reviewers to help improve the presentation of the paper.

\end{document}